# Windowed Decoding of Protograph-based LDPC Convolutional Codes over Erasure Channels

Aravind R. Iyengar, Marco Papaleo, Paul H. Siegel, *Fellow, IEEE*, Jack K. Wolf, *Life Fellow, IEEE*, Alessandro Vanelli-Coralli and Giovanni E. Corazza

***Abstract*—We consider a windowed decoding scheme for LDPC convolutional codes that is based on the belief-propagation (BP) algorithm. We discuss the advantages of this decoding scheme and identify certain characteristics of LDPC convolutional code ensembles that exhibit good performance with the windowed decoder. We will consider the performance of these ensembles and codes over erasure channels with and without memory. We show that the structure of LDPC convolutional code ensembles is suitable to obtain performance close to the theoretical limits over the memoryless erasure channel, both for the BP decoder and windowed decoding. However, the same structure imposes limitations on the performance over erasure channels with memory.**

## I. Introduction

LOW-DENSITY parity-check (LDPC) codes, although introduced in the early 1960's [4], were established as state-of-the-art codes only in the late 1990's with the application of statistical inference techniques [5] to graphical models representing these codes [6], [7]. The promising results from LDPC block codes encouraged the development of convolutional codes defined by sparse parity-check matrices.

LDPC convolutional codes (LDPC-CC) were first introduced in [8]. Ensembles of LDPC-CC have several attractive characteristics, such as thresholds approaching capacity with belief-propagation (BP) decoding [9], and BP thresholds close to the maximum a-posteriori (MAP) thresholds of random ensembles with the same degree distribution [10]. Whereas irregular LDPC block codes have also been shown to have BP thresholds close to capacity [11], the advantage with convolutional counterparts is that good performance is achieved by relatively simple regular ensembles. Also, the construction of finite-length codes from LDPC-CC ensembles can be readily optimized to ensure desirable properties, e.g. large girths and fewer cycles, using well-known techniques of LDPC code design. Most of these attractive features of LDPC-CC are pronounced when the blocklengths are large. However, BP decoding for these long codes might be computationally impractical. By implementing a *windowed decoder*, one can get around this problem.

In this paper, a windowed decoding scheme brought to the attention of the authors by Liva [12] is considered. This scheme exploits the convolutional structure of the parity-check matrix of the LDPC-CC to decode non-terminated codes, while maintaining many of the key advantages of iterative decoding schemes like the BP decoder, especially the low complexity and superior performance. Note that although similar decoding schemes were proposed in [13], [14], the aim in these papers was not to reduce the decoding latency or complexity. When used to decode terminated (block) LDPC-CC, the windowed decoder provides a simple, yet efficient way to trade-off decoding performance for reduced latency. Moreover, the proposed scheme provides the flexibility to set and change the decoding latency on the fly. This proves to be an extremely useful feature when the scheme is used to decode codes over upper layers of the internet protocol.

Our contributions in this paper are to study the requirements of LDPC-CC ensembles for good performance over erasure channels with windowed decoding (WD). We are interested in identifying characteristics of ensembles that present a good performance-latency trade-off. Further we seek to find such ensembles that are able to withstand not just random erasures but also long bursts of erasures. We reiterate that we will be interested in designing ensembles that have the aforementioned properties, rather than designing codes themselves. Although the channels considered here are erasure channels, we note that the WD scheme can be used when the transmission happens over any channel.

This paper is organized as follows. Section II introduces LDPC convolutional codes and the notation and terminology that will be used throughout the paper. In Section III we describe the decoding algorithms that will be considered. Along with a brief description of the belief-propagation algorithm, we will introduce the windowed decoding scheme that is based on BP. Possible variants of the scheme will also be discussed. Section IV deals with the performance of LDPC-CC on the binary erasure channel. Starting with a short recapitulation of known results for BP decoding, we will discuss the asymptotic analysis of the WD scheme in detail. Finite-length analysis will include performance evaluation using simulations that reinforce the observations made in the analysis. For erasure

A. R. Iyengar and P. H. Siegel are with the Department of Electrical and Computer Engineering and the Center for Magnetic Recording Research, University of California, San Diego, La Jolla, CA 92093 USA (e-mail: aravind@ucsd.edu, psiegel@ucsd.edu). J. K. Wolf (deceased) was with the Department of Electrical and Computer Engineering and the Center for Magnetic Recording Research, University of California, San Diego, La Jolla, CA 92093 USA. The work of A. R. Iyengar is supported by the National Science Foundation under the Grant CCF-0829865.

M. Papaleo is with Qualcomm Inc., San Diego, CA USA (e-mail: mpapaleo@qualcomm.com). A. Vanelli-Coralli and G. E. Corazza are with the University of Bologna, DEIS-ARCES, Viale Risorgimento, 2 - 40136 Bologna, Italy (e-mail: avanelli@arces.unibo.it, gecorazza@arces.unibo.it)

channels with memory, we analyse LDPC-CC ensembles both in the asymptotic setting and for finite lengths in Section V. We also include simulations illustrating the good performance of codes derived from the designed protographs over the Gilbert-Elliott channel. Finally, we summarize our findings in Section VI.

## II. LDPC Convolutional Codes

In the following, we will define LDPC-CC, give a construction starting from *protographs*, and discuss various ways of specifying ensembles of these codes.

### A. Definition

A rate $R = b/c$ binary, time-varying LDPC-CC is defined as the set of semi-infinite binary row vectors $\mathbf{v}_{[\infty]}$, satisfying $\mathbf{H}_{[\infty]}\mathbf{v}_{[\infty]}^{\mathrm{T}} = \mathbf{0}_{[\infty]}^{\mathrm{T}}$, where $\mathbf{H}_{[\infty]}$ is the *parity-check matrix*

$$\mathbf{H}_{[\infty]} = \begin{pmatrix} \mathbf{H}_0(1) & & & & \\ \mathbf{H}_1(1) & \mathbf{H}_0(2) & & & \\ \vdots & \mathbf{H}_1(2) & \ddots & & \\ \mathbf{H}_{m_s}(1) & \vdots & \ddots & \mathbf{H}_0(t) & \\ & \mathbf{H}_{m_s}(2) & \ddots & \mathbf{H}_1(t) & \ddots \\ & & \ddots & \vdots & \ddots \\ & & & \mathbf{H}_{m_s}(t) & \ddots \\ & & & & \ddots \end{pmatrix} \quad (1)$$

and $\mathbf{0}_{[\infty]}$ is the semi-infinite all-zero row vector. The elements $\mathbf{H}_i(t)$, $i = 0, 1, \cdots, m_s$ in (1) are binary matrices of size $(c-b) \times c$ that satisfy [15]

- $\mathbf{H}_i(t) = \mathbf{0}$, for $i < 0$ and $i > m_s$, $\forall\ t \geq 1$
- $\exists\ t > 0$ such that $\mathbf{H}_{m_s}(t) \neq \mathbf{0}$
- $\mathbf{H}_0(t)$ has full rank $\forall\ t \geq 1$.

The parameter $m_s$ is called the *memory* of the code and $\nu_s = (m_s + 1)c$ is referred to as the *constraint length*. The first two conditions above guarantee that the code has memory $m_s$ and the third condition ensures that the parity-check matrix is full-rank. In order to get sparse graph codes, the Hamming weight of each column $\mathbf{h}$ of $\mathbf{H}_{[\infty]}$ must be very low, i.e., $w_H(\mathbf{h}) \ll \nu_s$. Based on the matrices $\mathbf{H}_i(t)$, LDPC-CC can be classified as follows [8]. An LDPC-CC is said to be *periodic* if $\mathbf{H}_i(t) = \mathbf{H}_i(t+\tau)\ \forall\ i = 0, 1, \cdots, m_s,\ \forall\ t$ and for some $\tau > 1$. When $\tau = 1$, the LDPC-CC is said to be *time-invariant*, in which case the time dependence can be dropped from the notation, i.e. $\mathbf{H}_i(t) = \mathbf{H}_i\ \forall\ i = 0, 1, \cdots, m_s,\ \forall\ t$. If neither of these conditions holds, it is said to be *time-variant*.

Terminated LDPC-CC have a finite parity-check matrix

$$\mathbf{H}_{[L]} = \begin{pmatrix} \mathbf{H}_0(1) & & & \\ \mathbf{H}_1(1) & \mathbf{H}_0(2) & & \\ \vdots & \mathbf{H}_1(2) & \ddots & \\ \mathbf{H}_{m_s}(1) & \vdots & \ddots & \mathbf{H}_0(L) \\ & \mathbf{H}_{m_s}(2) & \ddots & \mathbf{H}_1(L) \\ & & \ddots & \vdots \\ & & & \mathbf{H}_{m_s}(L) \end{pmatrix}$$

where we say that the convolutional code has been terminated after $L$ *instants*. Such a code is said to be $(J, K)$ regular if $\mathbf{H}_{[L]}$ has exactly $J$ 1's in every column and $K$ 1's in every row excluding the first and the last $m_s(c-b)$ rows, i.e. ignoring the terminated portion of the code. It follows that for a given $J$, the parity-check matrix can be made sparse by increasing $c$ or $m_s$ or both, leading to different code constructions [16]. In this paper, we will consider LDPC-CC characterized by large $c$ and small $m_s$. As in [9], we will focus on regular LDPC-CC which can be constructed from a protograph.

### B. Protograph-based LDPC-CC

A protograph [17] is a relatively small bipartite graph from which a larger graph can be obtained by a copy-and-permute procedure—the protograph is copied $M$ times, and then the edges of the individual replicas are permuted among the $M$ replicas to obtain a single, large bipartite graph referred to as the *derived graph*. We will refer to $M$ as the *expansion factor*. $M$ is also referred to as the *lifting factor* in literature [11]. Suppose the protograph possesses $N_P$ variable nodes (VNs) and $M_P$ check nodes (CNs), with degrees $J_j, j = 1, \cdots, N_P$, and $K_i, i = 1, \cdots, M_P$, respectively. Then the derived graph will consist of $n = N_P M$ VNs and $m = M_P M$ CNs. The nodes of the protograph are labeled so that if the VN $V_j$ is connected to the CN $C_i$ in the protograph, then $V_j$ in a replica can only connect to one of the $M$ replicated $C_i$'s.

Protographs can be represented by means of an $M_P \times N_P$ bi-adjacency matrix $\mathbf{B}$, called the *base matrix* of the protograph where the entry $\mathbf{B}_{i,j}$ represents the number of edges between CN $C_i$ and VN $V_j$ (a non-negative integer, since parallel edges are permitted). The degrees of the VNs (CNs respectively) of the protograph are then equal to the sum of the corresponding column (row, respectively) of $\mathbf{B}$. A $(J, K)$ regular protograph-based code is then one with a base matrix where all VNs have degree $J$ and all CNs, excluding those in the terminated portion of the code, have degree $K$.

In terms of the base matrix, the copy-and-permute operation is equivalent to replacing each entry $\mathbf{B}_{i,j}$ in the base matrix with the sum of $\mathbf{B}_{i,j}$ distinct size-$M$ permutation matrices. This replacement is done ensuring that the degrees are maintained, e.g., a 2 in the matrix $\mathbf{B}$ is replaced by a matrix $\mathbf{H}_2^{(M)} = \mathbf{P}_1^{(M)} \oplus \mathbf{P}_2^{(M)}$ where $\mathbf{P}_1^{(M)}$ and $\mathbf{P}_2^{(M)}$ are two permutation matrices of size $M$ chosen to ensure that each row and column of $\mathbf{H}_2^{(M)}$ has two ones. The resulting matrix after the above transformation for each element of $\mathbf{B}$, which is the biadjacency matrix of the derived graph, corresponds to the parity-check matrix $\mathbf{H}$ of the code. The derived graph therefore is nothing but the *Tanner graph* corresponding to the parity-check matrix $\mathbf{H}$ of the code.

For different values of the expansion factor $M$, different blocklengths of the derived Tanner graph can be achieved, keeping the original graph structure imposed by the protograph. We can hence think of protographs as defining code ensembles that are themselves subsets of random LDPC code ensembles. We will henceforth refer to a protograph $\mathbf{B}$ and the ensemble $\mathcal{C}$ it represents interchangeably. This means that the density evolution analysis for the ensemble of codes represented by the protograph can be performed within the protograph. Furthermore, the structure imposed by a protograph on the derived graph can be exploited to design fast decoders and efficient encoders. Protographs give the code designer a refined control on the derived graph edge connections, facilitating good code design.

Analogous to LDPC block codes, LDPC-CC can also be derived by a protograph expansion. As for block codes, the parity-check matrices of these convolutional codes are composed of blocks of size-$M$ square matrices. We now give two constructions of $(J, K)$ regular LDPC-CC ensembles.

*1) **Classical construction**:* We briefly describe the construction introduced in [18]. For convenience, we will refer to this construction as the *classical* construction of $(J, K)$ regular LDPC-CC ensembles. Let $a$ be the greatest common divisor (gcd) of $J$ and $K$. Then there exist positive integers $J'$ and $K'$ such that $J = aJ'$, $K = aK'$, and $\gcd(J', K') = 1$. Assuming we terminate the convolutional code after $L$ instants, we obtain a block code, described by the base matrix

$$\mathbf{B}_{[L]} = \overbrace{\begin{pmatrix} \mathbf{B}_0 & & & \\ \mathbf{B}_1 & \mathbf{B}_0 & & \\ \vdots & \mathbf{B}_1 & \ddots & \\ \mathbf{B}_{m_s} & \vdots & \ddots & \mathbf{B}_0 \\ & \mathbf{B}_{m_s} & \ddots & \mathbf{B}_1 \\ & & \ddots & \vdots \\ & & & \mathbf{B}_{m_s} \end{pmatrix}}^{L}$$

where $m_s = a-1$ is the *memory* of the LDPC-CC and $\mathbf{B}_i, i = 0, \cdots, m_s$ are $J' \times K'$ submatrices that are all identical and have all entries equal to 1. Note that an LDPC-CC constructed from the protograph with base matrix $\mathbf{B}_{[L]}$ could be time-varying or not depending on the expansion of the protograph into the parity-check matrix.

The protograph of the terminated code has $N_P = LK'$ VNs and $M_P = (L + m_s)J'$ CNs. The rate of the LDPC-CC is therefore

$$R_L = 1 - \left(\frac{L + m_s}{L}\right)\frac{J'}{K'} = 1 - \left(1 + \frac{m_s}{L}\right)(1 - R) \quad (2)$$

where $R = 1 - \frac{J'}{K'}$ is the rate of the non-terminated code. Note that $R_L \to R$ and the LDPC-CC has a regular degree distribution [9] when $L \to \infty$. We will assume that the parameters satisfy $K' > J'$ and $L \geq \frac{1-R}{R} m_s$ so that the rates $R$ and $R_L$ of the non-terminated and terminated codes, respectively, are in the proper range.

The classical construction was proposed in [18] and it produces protographs for some $(J, K)$ regular LDPC-CC ensembles. However, not all $(J, K)$ regular LDPC-CC can be constructed, e.g. $m_s$ becomes zero if $J$ and $K$ are relatively prime and consequently the resulting code has no memory. In [9], the authors addressed this problem by proposing a construction rule based on *edge spreading*. We denote an ensemble of $(J, K)$ regular LDPC-CC constructed as described here as $\mathcal{C}_c(J, K)$ with the subscript $c$ for "classical" construction.

*2) **Modified construction**:* We propose a modified construction that is similar to the classical construction except that we do not require that $m_s = a - 1$, i.e. the memory of the LDPC-CC is independent of its degree distribution. We further disregard the requirement that the $\mathbf{B}_i$ matrices are identical and have only ones, i.e. parallel edges in the protograph are allowed. However, the sizes of the submatrices $\mathbf{B}_i, i = 0, 1, \cdots, m_s$ will still be $J' \times K'$. We will denote a $(J, K)$ regular LDPC-CC ensemble constructed in this manner as $\mathcal{C}_m(J, K)$, with subscript $m$ for "modified" construction. Note that the rate of the $\mathcal{C}_m(J, K)$ ensemble is still given by Equation (2). Further, the independence of the code memory and the degree distribution allows us to construct LDPC-CC even when $J$ and $K$ are co-primes. This is illustrated in the following example.

*Example 1:* Let $J = 3$ and $K = 4$. Clearly, a classical construction of this ensemble is not possible. However, with the modified construction, we can set $m_s = 1$ and define the ensemble $\mathcal{C}_m(J, K)$ given by

$$\mathbf{B}_0 = \begin{pmatrix} 1 & 0 & 1 & 1 \\ 0 & 1 & 0 & 1 \\ 1 & 1 & 1 & 0 \end{pmatrix}, \mathbf{B}_1 = \begin{pmatrix} 1 & 0 & 0 & 0 \\ 0 & 1 & 0 & 1 \\ 0 & 0 & 1 & 0 \end{pmatrix}$$

with design rate $R_L = 1 - \frac{3}{4}\left(\frac{L+1}{L}\right)$ for a termination length $L$. Note that these submatrices are by no means the only possible ones. Another set of submatrices satisfying the constraints is

$$\hat{\mathbf{B}}_0 = \begin{pmatrix} 2 & 0 & 0 & 1 \\ 0 & 2 & 0 & 1 \\ 0 & 0 & 2 & 0 \end{pmatrix}, \hat{\mathbf{B}}_1 = \begin{pmatrix} 0 & 0 & 1 & 0 \\ 0 & 0 & 0 & 1 \\ 1 & 1 & 0 & 0 \end{pmatrix}.$$

□

The above example brings out the similarity between the proposed modified construction and the technique of edge spreading employed in [9], wherein the edges of the protograph defined by the matrix

$$\mathbf{B}'_0 = \begin{pmatrix} 2 & 0 & 1 & 1 \\ 0 & 2 & 0 & 2 \\ 1 & 1 & 2 & 0 \end{pmatrix}$$

are "spread" between the matrices $\mathbf{B}_0$ and $\mathbf{B}_1$ (or between $\hat{\mathbf{B}}_0$ and $\hat{\mathbf{B}}_1$) to obtain a $(3, 4)$ regular LDPC-CC ensemble with memory $m_s = 1$. The advantage of the modified construction is thus clear—it gives us more degrees of freedom to design the protographs in comparison with the classical construction. In particular, the ensemble specified by the classical construction is contained in the set of ensembles allowed by the modified construction, meaning that the best performing $\mathcal{C}_m(J, K)$ ensemble (with memory the same as that of the $\mathcal{C}_c(J, K)$ ensemble) is at least as good as the $\mathcal{C}_c(J, K)$ ensemble. Note that in [9], there was no indication as to how edges are to be spread between matrices. With windowed decoding, we will shortly show that different protographs (edge spreadings) have different performances. We will also identify certain design criteria for efficient modified constructions that suit windowed decoding.

### C. Polynomial representation of LDPC-CC ensembles

We have thus far specified LDPC-CC ensembles by giving the parameter $L$ and the matrices $\mathbf{B}_i, i = 0, 1, \cdots, m_s$. An alternative specification of terminated protograph-based LDPC-CC ensembles using polynomials is useful in establishing certain properties of $(J, K)$ regular ensembles and is described below.

Instead of specifying $(m_s+1)$ matrices $\mathbf{B}_i$ of size $J' \times K'$, we can specify the $K'$ columns of the $(m_s+1)J' \times K'$ matrix

$$\mathbf{B}_{[1]} = \begin{pmatrix} \mathbf{B}_0 \\ \mathbf{B}_1 \\ \vdots \\ \mathbf{B}_{m_s} \end{pmatrix}$$

using a polynomial of degree no more than $\mathtt{d} = (m_s+1)J' - 1$ for each column. The polynomial of the $j^{\text{th}}$ column

$$p_j(x) = p_j^{(0)} + p_j^{(1)}x + p_j^{(2)}x^2 + \cdots + p_j^{(\mathtt{d})}x^{\mathtt{d}} \tag{3}$$

is defined so that the coefficient of $x^i$, $p_j^{(i)}$, is the $(i+1, j)$ entry of $\mathbf{B}_{[1]}$ for all $i = 0, 1, \cdots, \mathtt{d}$ and $j = 1, 2, \cdots, K'$. Therefore, an equivalent way of specifying the LDPC-CC ensemble is by giving $L$ and the set of polynomials $\{p_j(x), j = 1, 2, \cdots, K'\}$. With this notation, the $l^{\text{th}}$ column of $\mathbf{B}_{[L]}$ is specified by the polynomial $x^{J'i}p_j(x)$ where $l = iK' + j$ for unique $0 \le i \le L-1$ and $1 \le j \le K'$. We can hence use "the column index" and "the column polynomial" interchangeably. Further, to define $(J, K)$ regular ensembles, we will need the constraints

$$p_j(1) = J \ \forall\ 1 \le j \le K'$$

and

$$\sum_{j=1}^{K'} p_j^{[m]}(1) = K \ \forall\ 0 \le m \le J' - 1,$$

where $p_j^{[m]}(x)$ is the polynomial of degree no larger than $m_s$ obtained from $p_j(x)$ by collecting the coefficients of terms with degrees $l$ where $l = hJ' + m$ for some $0 \le h \le m_s$, i.e. $l = m (\text{mod } J')$:

$$\begin{aligned} p_j^{[m]}(x) &= p_j^{(m)} + p_j^{(J'+m)}x + \cdots + p_j^{(m_sJ'+m)}x^{m_s} \\ &= \sum_{h=0}^{m_s} p_j^{(hJ'+m)}x^h. \end{aligned} \tag{4}$$

We will refer to these polynomials as the *modulo polynomials*. Let us denote the set of polynomials defining an LDPC-CC ensemble as $\mathcal{P} = \{p_j(x), j \in [K']\}$, where $[K'] = \{1, 2, \cdots, K'\}$, and the modulo polynomials as $\mathcal{P}_l = \{p_j^{[l]}(x), j \in [K']\}$, $l = 0, 1, \cdots, J'-1$. Later in the paper, we will say "the summation of polynomials $p_i(x)$ and $p_j(x)$" to mean the collection of the $i^{\text{th}}$ and the $j^{\text{th}}$ columns of $\mathbf{B}_{[1]}$. The following example illustrates the notation.

*Example 2:* For $(J, 2J)$ codes, we have $J' = 1$ and $K' = 2$, the component base matrices $\mathbf{B}_i, i = 0, ..., m_s$ are $1 \times 2$ matrices. With the first column of the protograph $\mathbf{B}_{[1]}$, we associate a polynomial $p_1(x) = p_1^{(0)} + p_1^{(1)}x + \cdots + p_1^{(m_s)}x^{m_s}$ of degree at most $m_s$. Similarly, with the second column we associate a polynomial $p_2(x) = p_2^{(0)} + p_2^{(1)}x + \cdots + p_2^{(m_s)}x^{m_s}$, also of degree at most $m_s$. Then, the $(2i+1)^{\text{th}}$ column of $\mathbf{B}_{[L]}$ can be associated with the polynomial $x^ip_1(x)$, and the $(2i+2)^{\text{th}}$ column with the polynomial $x^ip_2(x)$. As noted earlier, we will use the polynomial of a column and its index interchangeably, e.g. when we say "choosing the polynomial $x^ip_1(x)$," we mean that we choose the $(2i+1)^{\text{th}}$ column of $\mathbf{B}_{[L]}$. Similarly, by "summations of polynomials $p_1(x)$ and $p_2(x)$," we mean the collection of the corresponding columns of $\mathbf{B}_{[L]}$. In order to define $(J, 2J)$ regular ensembles, we will further have the constraint $p_1(1) = p_2(1) = J$. In this case, since $J' = 1$, $p_1^{[0]}(1) + p_2^{[0]}(1) = 2J$ is the same as the previous constraint, because $p_1^{[0]}(1) + p_2^{[0]}(1) = p_1(1) + p_2(1)$. □

We define the *minimum degree* of a polynomial $a(x)$ as the least exponent of $x$ with a positive coefficient and denote it as $\min\deg(a(x))$. Clearly, $0 \le \min\deg(a(x)) \le \deg(a(x))$. Let us define a partial ordering of polynomials with non-negative integer coefficients as follows. We write $a(x) \preceq b(x)$ if $\min\deg(a(x)) = \min\deg(b(x))$, $\deg(a(x)) = \deg(b(x))$ and the coefficients of $a(x)$ are no larger than the corresponding ones of $b(x)$. The ordering $\preceq$ satisfies the following properties over polynomials with non-negative integer coefficients: if $a(x) \preceq b(x)$ and $c(x) \preceq d(x)$, then

$$\begin{aligned} a(x) + c(x) &\preceq b(x) + d(x) \\ a(x)c(x) &\preceq b(x)d(x). \end{aligned}$$

We define the *boundary polynomial* $\beta(a(x))$ of a polynomial $a(x)$ to be $\beta(a(x)) = x^i + x^j$ where $i = \min\deg(a(x))$ and $j = \deg(a(x))$. Note that when $i = j$, we define $\beta(a(x)) = x^i$. We have for any polynomial $a(x)$, $\beta(a(x)) \preceq a(x)$.

## III. Decoding Algorithms

LDPC-CC are characterized by a very large constraint length $\nu_s = (m_s+1)K'M$. Since the Viterbi decoder has a complexity that scales exponentially in the constraint length, it is impractical for this kind of code. However, the sparsity of the parity-check matrix can be exploited and an iterative message passing algorithm can be adopted for decoding. We consider two specific iterative decoders here—a conventional belief-propagation decoder [6], [19] and a variant called a windowed decoder.

### A. Belief-Propagation (BP)

For terminated LDPC-CC, decoding can be performed as in the case of an LDPC block code, meaning that each frame carrying a codeword obtained through the termination can be decoded with the sum-product algorithm (SPA) [19].

Note that since the BP decoder can start decoding only after the entire codeword is received, the total decoding latency $\Lambda_{BP}$ is given by $\Lambda_{BP} = T_{cw} + T_{dec}$, where $T_{cw}$ is the time taken to receive the entire codeword and $T_{dec}$ is the time needed to decode the codeword. In many practical applications this latency is large and undesirable. Moreover, for non-terminated LDPC-CC, a BP decoder cannot be employed.

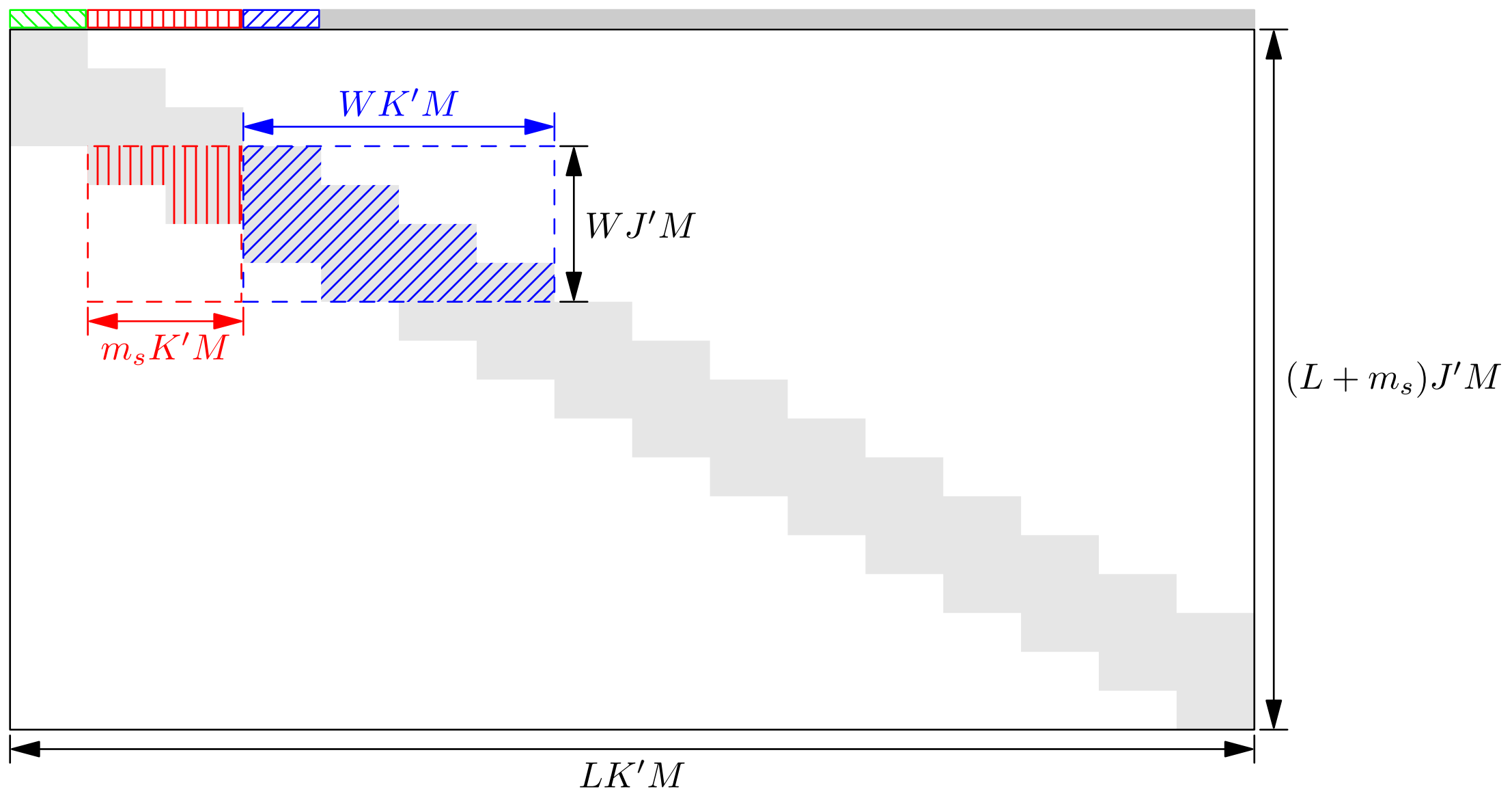

Fig. 1. Illustration of windowed decoding (WD) with window of size $W = 4$ for a $\mathcal{C}_m(J, 2J)$ LDPC-CC with $m_s = 2$ and $L = 16$ at the fourth decoding instant. This window configuration consists of $J_W = WJ'M = 4M$ rows of the parity-check matrix and all the $(W + m_s)K'M = 12M$ columns involved in these equations: this comprises the red (vertically hatched) and the blue (hatched) edges shown within the matrix. Note that the symbols shown in green (backhatched) above the parity-check matrix have all been processed. The targeted symbols are shown in blue (hatched) above the parity-check matrix and the symbols that are yet to be decoded are shown in gray above the parity-check matrix.

### B. *Windowed Decoding (WD)*

The convolutional structure of the code imposes a constraint on the VNs connected to the same parity-check equations—two VNs of the protograph that are at least $(m_s + 1)K'$ columns apart cannot be involved in the same parity-check equation. This characteristic can be exploited in order to perform continuous decoding of the received stream through a "window" that slides along the bit sequence. Moreover, this structure allows for the possiblity of parallelizing the iterations of the message passing decoder through several processors working in different regions of the Tanner graph. A pipeline decoder based on this idea was proposed in [8]. In this paper we consider a windowed decoder to decode terminated codes with reduced latency. Note that whereas a similar sliding window decoder was used to bound the performance of BP decoding in [14], we are interested in evaluating the performance of the windowed decoder from a perspective of reducing the decoding complexity and latency.

Consider a terminated $(J, K)$ regular parity-check matrix $\mathbf{H}$ built from a base matrix $\mathbf{B}$. The windowed decoder works on sub-protographs of the code and the window size $W$ is defined as the number of sets of $J'$ CNs of the protograph $\mathbf{B}$ considered within each window. In the parity-check matrix $\mathbf{H}$, the window thus consists of $J_W = WJ'M = W(c - b)$ rows of $\mathbf{H}$ and all columns that are involved in the check equations corresponding to these rows. We will henceforth refer to the size of the window only in terms of the protograph with the corresponding size in the parity-check matrix implied. The window size $W$ ranges between $(m_s+1)$ and $(L-1)$ because each VN in the protograph is involved in at most $J'(m_s + 1)$ check equations; and, although there are a total of $M_P = J'(L + m_s)$ CNs in $\mathbf{B}$, the decoder can perform BP when all the VN symbols are received, i.e. when $L \leq W \leq L + m_s$. Apart from the window size, the decoder also has a (typically small) target erasure probability $\delta \geq 0$ as a parameter[1]. The aim of the WD is to reduce the erasure probability of every symbol in the codeword to a value no larger than $\delta$.

At the first decoding instant, the decoder performs belief-propagation over the edges within the window with the aim of decoding all of the first $K'$ symbols in the window, called the *targeted symbols*. The window slides down $J'$ rows and right $K'$ columns in $\mathbf{B}$ after at least a fraction $(1 - \delta)$ of the targeted symbols are recovered (or, in general, after a maximum number of belief-propagation iterations have been performed), and continues decoding at the new position at the next decoding time instant.

We refer to the set of edges included in the window at any particular decoding time instant as the *window configuration*. In the terminated portion of the code, the window configuration will have fewer edges than other configurations within the code. Since the WD aims to recover only the targeted symbols within each window configuration, the entire codeword is recovered in $L$ decoding time instants. Fig. 1 shows a schematic representation of the WD for $W = 4$.

The decoding latency of the $K'$ targeted symbols with WD is therefore given by $\Lambda_{WD} = T_W + T_{dec}(W)$, where $T_W$ is the time taken to receive all the symbols required to decode the $K'$ targeted symbols, and $T_{dec}(W)$ is the time taken to decode the targeted symbols. The parameters $T_{cw}$ and $T_W$ are related as

$$T_W = \frac{(W + m_s)K'}{LK'} T_{cw} = \frac{W + m_s}{L} T_{cw},$$

since at most $(W + m_s)K'$ symbols are to be received to process the targeted symbols. The relation between $T_{dec}$ and

[1]We will see shortly that setting $\delta = 0$ is not necessarily the most efficient use of the WD scheme.

$T_{dec}(W)$ is given by

$$T_{dec}(W) = \frac{W}{L} T_{dec},$$

since the complexity of BP decoding scales linearly in blocklength and the WD uses BP decoding over $WK'$ symbols in each window configuration. We assume that the number of iterations of message passing performed is fixed to be the same for the BP decoder and the WD. Thus, in latency-limited scenarios, we can use the WD to obtain a latency reduction of

$$\Lambda_{WD} \leq \frac{W + m_s}{L} \Lambda_{BP} \triangleq w \Lambda_{BP}.$$

The smallest latency supported by the code-decoder system is therefore at most a fraction $w_{min} = \frac{2m_s+1}{L}$ that of the BP decoder. As pointed out earlier, the only choice for non-terminated codes is to use some sort of a windowed decoder. For the sequence of ensembles indexed by $L$, with the choice of the proposed WD with a fixed finite window size $W$, the decoding latency vanishes as $O(\frac{1}{L})$. We will typically be interested in small values of $W$ where large gains in decoding latencies are achievable. Since the decoding latency increases as $W$ increases, the trade-off between decoding performance and latency can be studied by analyzing the performance of the WD for the entire range of window sizes.

*Latency Flexibility:* Although reduced latency is an important characteristic of WD, what is perhaps more useful practically is the flexibility to alter the latency with suitable changes in the code performance. The latency can be controlled by varying the parameter $W$ as required. If a large latency can be handled, $W$ can be kept large ensuring good code performance and if a small latency is required, $W$ can be made small while paying a price with the code performance (We will see shortly that the performance of WD is monotonic in the window size).

One possible variant of WD is a decoding scheme which starts with the smallest possible window size and the size is increased whenever targeted symbols cannot be decoded, i.e., the target erasure probability cannot be met within the fixed maximum number of iterations. Other schemes where the window size is either increased or decreased based on the performance of the last few window configurations are also possible.

## IV. Memoryless Erasure Channels

In this section, we confine our attention to the performance of the LDPC-CC when the transmission occurs over a memoryless erasure channel, i.e. a binary erasure channel (BEC) parameterized by the channel erasure rate $\varepsilon$.

### A. Asymptotic analysis

We consider the performance of the LDPC-CC in terms of the average performance of the codes belonging to ensembles defined by protographs in the limit of infinite blocklengths and in the limit of infinite iterations of the decoder. As in the case of LDPC block codes, the ensemble average performance is a good estimate of the performance of a code in the ensemble with high probability. We will therefore concentrate on the erasure rate *thresholds* [11] of the code ensembles as a performance metric in our search for good LDPC-CC ensembles.

*1) **BP**:* The asymptotic analysis of LDPC block codes with the BP decoder over the BEC has been well studied [20]–[23]. For LDPC-CC based on protographs, the BP decoding thresholds can be numerically estimated using the Protograph-EXIT (P-EXIT) analysis [24]. This method is similar to the standard EXIT analysis in that it tracks the mutual information between the message on an edge and the bit value corresponding to the VN on which the edge is incident, while maintaining the graph structure dictated by the protograph[2].

The processing at a CN of degree $d_C$ results in an updating of the mutual information on the $d_C^{\text{th}}$ edge as

$$I_{\text{out},d_C} = \mathtt{C}\left(I_{\text{in},1}, \cdots, I_{\text{in},d_C-1}\right) = \prod_{i=1}^{d_C-1} I_{\text{in},i} \tag{5}$$

and the corresponding update at a VN of degree $d_V$ gives

$$I_{\text{out},d_V} = \mathtt{V}\left(I_{ch}, I_{\text{in},1}, \cdots, I_{\text{in},d_V-1}\right) = 1 - \varepsilon \prod_{i=1}^{d_V-1} \left(1 - I_{\text{in},i}\right) \tag{6}$$

where $I_{ch} = 1-\varepsilon$ is the mutual information obtained from the channel. Note that the edge multiplicities are included in the above check and variable node computations. The a-posteriori mutual information $\mathcal{I}$ at a VN is found using

$$\mathcal{I} = 1 - \varepsilon \prod_{i=1}^{d_V} (1 - I_{\text{in},i}) = 1 - (1 - I_{\text{out},d_V})(1 - I_{\text{in},d_V})$$

where the second equality follows from (6). The decoder is said to be successful when the a-posteriori mutual information at all the VNs of the protograph converges to 1 as the number of iterations of message passing goes to infinity. The *BP threshold* $\varepsilon^*_{BP}(\mathbf{B})$ of the ensemble described by the protograph with base matrix $\mathbf{B}$ is defined as the supremum of all erasure rates for which the decoder is successful.

*Example 3:* The protograph $\mathbf{B}_{3,6} = (3 \quad 3)$ has a BP threshold of $\varepsilon^*_{BP} \approx 0.4294$. Note that all the CNs in the protograph are of degree 6 while all the VNs are of degree 3. This BP threshold is expected because $\mathbf{B}_{3,6}$ corresponds to the $(3, 6)$ regular LDPC block code ensemble. The following protograph $\mathbf{B}'_{3,6}$ has a BP threshold $\varepsilon^*_{BP} \approx 0.4879$ for $L = 40$. Note that, as before, all VNs are of degree 3 and all the CNs except the ones in the terminated portion of the code are of degree 6.

[2]We will use the phrase "mutual information on an edge" to mean the mutual information between the message on the edge and the bit corresponding to the adjacent VN.

$$\mathbf{B}'_{3,6} = \begin{pmatrix} 1 & 1 & 0 & 0 & 0 & 0 & \cdots & 0 & 0 \\ 1 & 1 & 1 & 1 & 0 & 0 & \cdots & \vdots & \vdots \\ 1 & 1 & 1 & 1 & 1 & 1 & \cdots & 0 & 0 \\ 0 & 0 & 1 & 1 & 1 & 1 & \cdots & 0 & 0 \\ 0 & 0 & 0 & 0 & 1 & 1 & \cdots & 0 & 0 \\ 0 & 0 & 0 & 0 & 0 & 0 & \cdots & 1 & 1 \\ \vdots & \vdots & \vdots & \vdots & \vdots & \vdots & \ddots & 1 & 1 \\ 0 & 0 & 0 & 0 & 0 & 0 & \cdots & 1 & 1 \end{pmatrix}$$

This is the $\mathcal{C}_c(3,6)$ ensemble constructed in [18]. In terms of the notation introduced, this is given as $\mathbf{B}_0 = \mathbf{B}_1 = \mathbf{B}_2 = [1 \;\; 1]$; or equivalently as $p_1(x) = p_2(x) = 1 + x + x^2$. □

The above example illustrates the strength of protographs—they allow us to choose structures within an ensemble defined by a pair of degree distributions that may perform better than the ensemble average. In fact, the BP performance of regular LDPC-CC ensembles has been related to the *maximum-a-posteriori* (MAP) decoder performance of the corresponding unstructured ensemble [10].

*2) **WD**:* We now analyze the performance of the WD described in Section III-B in the limit of infinite blocklengths and the limit of infinite iterations of belief-propagation within each window.

*Remark:* In the limit of infinite blocklength, each term in the base protograph $\mathbf{B}$ is replaced by a permutation matrix of infinite size to obtain the parity-check matrix, and therefore the latency of any window size is infinite, apparently defeating the purpose of WD. Our interest in the asymptotic performance, however, is justified as it allows us to establish lower bounds on the probability of failure of the windowed decoder to recover the symbols of the finite length code. In practice, it is to be expected that the gap between the performance of a finite length code with WD and the asymptotic ensemble performance of the ensemble to which the code belongs increases as the window size reduces due to the reduction in the blocklength of the subcode defined by the window.

The asymptotic analysis for WD is very similar to that of the BP decoder owing to the fact that the part of the code within a window is itself a protograph-based code. However, the main distinction in this case is the definition of decoding success. In the case of BP decoding, the decoding is considered a success only when, for any symbol in the codeword, the probability of failing to recover the symbol goes to 0 (or equivalently, the a-posteriori mutual information goes to 1) as the number of rounds of message-passing goes to infinity. On the other hand, the decoding within a window is successful as long as the probability of failing to recover the targeted symbols becomes smaller than a predecided small value $\delta$. The decoder performance therefore depends on two parameters: the window size $W$ and the target erasure probability $\delta$.

We define the threshold $\varepsilon^*_{(i)}(\mathbf{B}, W, \delta)$ of the $i^{\text{th}}$ window configuration to be the supremum of the channel erasure rates for which the WD succeeds in retrieving the targeted symbols of the $i^{\text{th}}$ window with a probability at least $(1-\delta)$, given that each of the targeted symbols corresponding to the first $(i-1)$ window configurations is known with probability $1-\delta$. Fig. 2 illustrates the threshold $\varepsilon^*_{(i)}(\mathbf{B}, W, \delta)$ of the $i^{\text{th}}$ window configuration. The *windowed threshold* $\varepsilon^*(\mathbf{B}, W, \delta)$ is then defined as the supremum of channel erasure rates for which the windowed decoder can decode each symbol in the codeword with probability at least $1-\delta$.

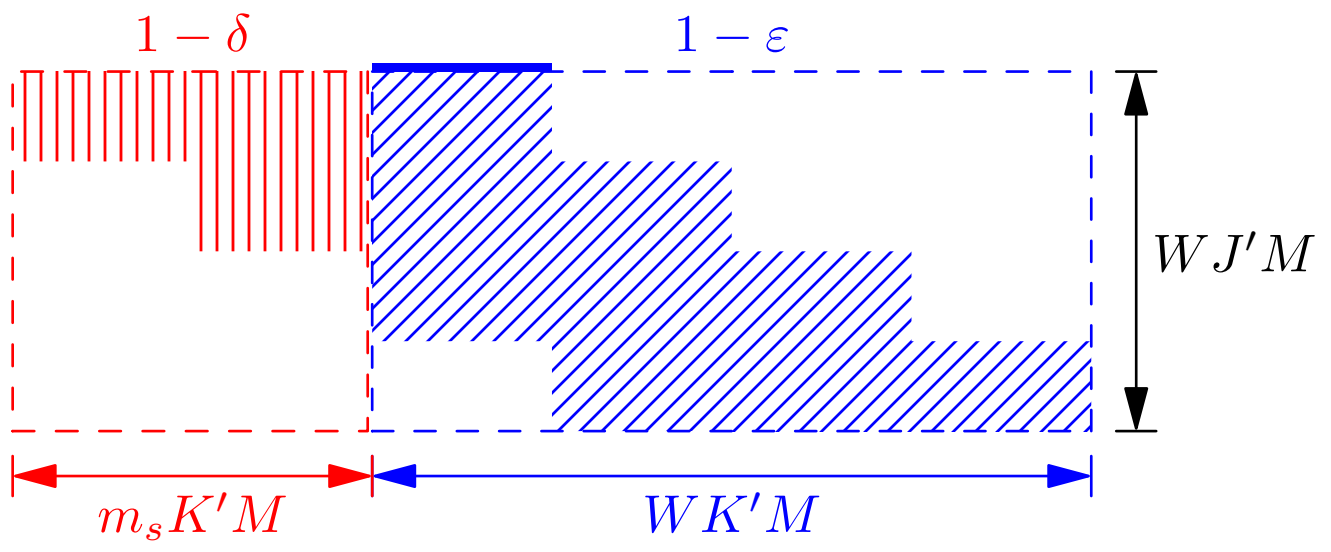

Fig. 2. Illustration of the threshold of the $i^{\text{th}}$ window configuration $\varepsilon^*_{(i)}(\mathbf{B}, W, \delta)$. The targeted symbols of the previous window configurations are known with probability $1-\delta$. The targeted symbols within the window are highlighted with a solid blue bar on top of the window. The symbols within the blue (hatched) region in the window are initially known with probability $1-\varepsilon$. The task of the decoder is to perform BP within this window until the erasure probability of the targeted symbols is smaller than $\delta$. The window is then slid to the next configuration.

We assume that between decoding time instants, no information apart from the targeted symbols is carried forward, i.e. when a particular window configuration has been decoded, all the present processing information apart from the decoded targeted symbols themselves is discarded. With this assumption, it is clear that the windowed threshold of a protograph-based LDPC-CC ensemble is given by the minimum of the thresholds of its window configurations. For the classical and modified constructions of LDPC-CC described in Section II-B, all window configurations are similar except the ones at the terminated portion of the code. Since the window configurations at the terminated portions can only perform better, the windowed threshold is determined by the threshold of a window configuration not in the terminated portion of the code. Note that the performance of WD when the information from processing the previous window configurations is made use of in successive window configurations, e.g. when symbols other than the targeted symbols that were decoded previously are also retained, can only be better than what we obtain here.

We now state a monotonicity property of the WD the proof of which is relegated to Appendix I.

*Proposition 1 (Monotonicity of WD performance in $W$):* For any $\mathcal{C}_m(J, K)$ ensemble $\mathbf{B}$,

$$\varepsilon^*(\mathbf{B}, W, \delta) \leq \varepsilon^*(\mathbf{B}, W+1, \delta).$$

■

It follows immediately from the definition of the windowed threshold that

$$\varepsilon^*(\mathbf{B}, W, \delta) \leq \varepsilon^*(\mathbf{B}, W, \delta') \;\forall\; \delta \leq \delta'.$$

Furthermore, from the continuity of the density evolution equations (6) and (5), we have that when we set $\delta = 0$, we decode not only the targeted symbols within the window but all the remaining symbols also. Since the symbols in the right end of the window are the "worst protected" ones within the window (in the sense that these are the symbols for which the least number of constraints are used to decode), we expect the windowed thresholds $\varepsilon^*(\mathbf{B}, W, \delta = 0)$ to be dictated mostly by the behavior of the submatrix $\mathbf{B}_0$ under BP. In the following, when the base matrix $\mathbf{B}$ of the protograph corresponding to an ensemble $\mathcal{C}$ is unambiguous, we will write $\varepsilon^*(\mathbf{B}, W, \delta)$ and $\varepsilon^*(\mathcal{C}, W, \delta)$ interchangeably.

We next turn to giving some properties of LDPC-CC ensembles with good performance under WD. We start with an example that illustrates the stark change in performance a small difference in the structure of the protograph can produce.

*Example 4:* Consider WD with the ensemble $\mathcal{C}_c(3, 6)$ in Example 3 with a window of size $W = 3$. The corresponding protograph defining the first window configuration is

$$\begin{pmatrix} 1 & 1 & 0 & 0 & 0 & 0 \\ 1 & 1 & 1 & 1 & 0 & 0 \\ 1 & 1 & 1 & 1 & 1 & 1 \end{pmatrix}$$

and we have $\varepsilon^*(\mathcal{C}_c(3, 6), W = 3, \delta = 0) = 0$. This is seen readily by observing that there are VNs of degree 1 that are connected to the same CNs. In fact, from this reasoning, we see that $\varepsilon^*(\mathcal{C}_c(J, K'J), W, \delta = 0) = 0 \ \forall \ J \leq W \leq L$.

As an alternative, we consider the modified construction of Section II-B2 to obtain the $\mathcal{C}_m(J, K)$ ensemble $\mathbf{B}'$ given by $\mathbf{B}_0 = [2 \ \ 2], \mathbf{B}_1 = [1 \ \ 1]$. This ensemble has a BP threshold $\varepsilon^*_{BP}(\mathbf{B}') \approx 0.4875$ for $L = 40$ which is quite close to that of the ensemble $\mathcal{C}_c(3, 6)$, $\varepsilon^*_{BP}(\mathcal{C}_c(3, 6)) \approx 0.4879$. WD with a window of size 3 for this ensemble has the first window configuration

$$\begin{pmatrix} 2 & 2 & 0 & 0 & 0 & 0 \\ 1 & 1 & 2 & 2 & 0 & 0 \\ 0 & 0 & 1 & 1 & 2 & 2 \end{pmatrix}$$

which has a threshold $\varepsilon^*(\mathbf{B}', W = 3, \delta = 0) \approx 0.3331$, i.e. we can theoretically get close to 68.3% of the BP threshold with $< 10\%$ of the latency of the BP decoder. Note that this improvement in threshold has been obtained while also increasing the rate of the ensemble, since $m_s = 1$ for the $\mathbf{B}'$ ensemble in comparison with $m_s = 2$ for $\mathcal{C}_c(3, 6)$. □

The above example illustrates the tremendous advantage obtained by using $\mathcal{C}_m(J, K)$ ensembles for WD even under the severe requirement of $\delta = 0$. The following is a good rule of thumb for constructing LDPC-CC ensembles that have good performance with WD.

> *Design Rule 1:* For $\mathcal{C}_m(J, K'J)$ ensembles, set $p_j^{(d_j)} \geq 2$ for all $j \in [K']$ where $d_j = \min \deg(p_j(x))$.

The above design rule says that for $(J, K'J)$ ensembles, it is better to avoid degree-1 VNs within a window. Note that none of the $\mathcal{C}_c(J, K'J)$ ensembles satisfy this design rule. We now illustrate the performance of LDPC-CC ensembles with WD when we allow $\delta > 0$.

*Example 5:* We compare three LDPC-CC ensembles. The first is the classical LDPC-CC ensemble $\mathcal{C}_1 = \mathcal{C}_c(3, 6)$. The second and the third are LDPC-CC ensembles constructed as described in Section II-B2. The ensemble $\mathcal{C}_2$ is defined by the polynomials

$$p_1(x) = 2 + x^2, p_2(x) = 2 + x$$

and $\mathcal{C}_3$ is defined by

$$q_1(x) = q_2(x) = 2 + x.$$

We first observe that all three ensembles have the same asymptotic degree distribution, i.e. all are $(3, 6)$ regular LDPC-CC ensembles when $L \to \infty$. While $\mathcal{C}_1$ and $\mathcal{C}_2$ have a memory $m_s = 2$, $\mathcal{C}_3$ has a memory $m_s = 1$. Therefore, for a fixed $L$, while $\mathcal{C}_1$ and $\mathcal{C}_2$ have the same rate, $\mathcal{C}_3$ has a higher rate. Another consequence of a smaller $m_s$ is that $\mathcal{C}_3$ can be decoded with a window of size $W_{min}(\mathcal{C}_3) = 2$. Further note that whereas $\mathcal{C}_2$ and $\mathcal{C}_3$ satisfy Design Rule 1, $\mathcal{C}_1$ does not. For a window of size 3, the subprotographs for ensembles $\mathcal{C}_1$ and $\mathcal{C}_3$ are as shown in Example 4, and that for ensemble $\mathcal{C}_2$ is as shown below

$$\begin{pmatrix} 2 & 2 & 0 & 0 & 0 & 0 \\ 0 & 1 & 2 & 2 & 0 & 0 \\ 1 & 0 & 0 & 1 & 2 & 2 \end{pmatrix}$$

In Fig. 3, we show the windowed thresholds plotted against the window size for the three ensembles $\mathcal{C}_1, \mathcal{C}_2$ and $\mathcal{C}_3$ by fixing $L = 100$ for $\delta \in \{10^{-6}, 10^{-12}\}$.

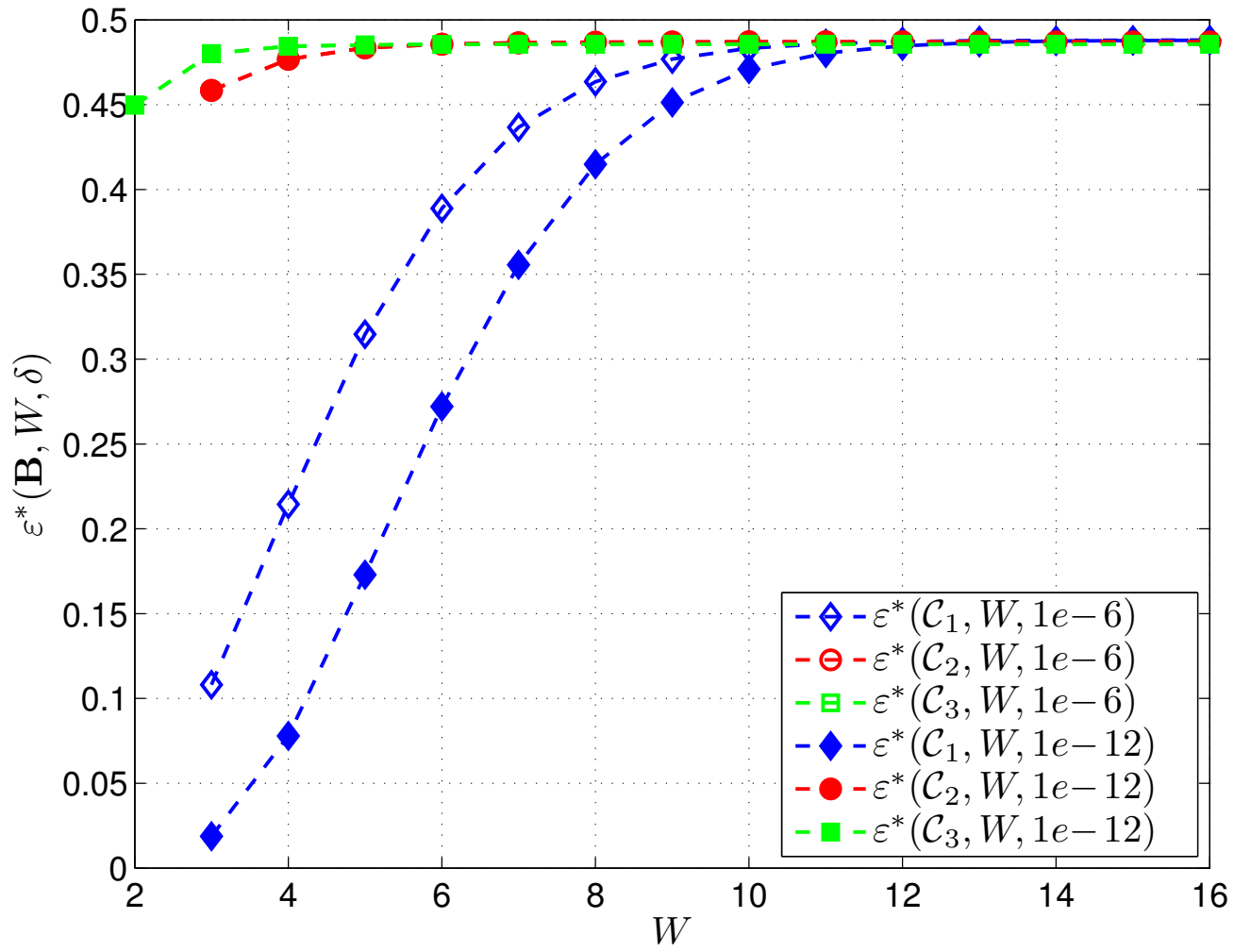

Fig. 3. Windowed threshold as a function of the window size for the ensembles $\mathcal{C}_i, i = 1, 2, 3$ with $\delta \in \{10^{-6}, 10^{-12}\}$. The rates of the ensembles $\mathcal{C}_1$ and $\mathcal{C}_2$ are 0.49 whereas that of $\mathcal{C}_3$ is 0.495. The corresponding Shannon limits are therefore 0.51 for $\mathcal{C}_1$ and $\mathcal{C}_2$, and 0.505 for $\mathcal{C}_3$.

A few observations are in order. The monotonicity of $\varepsilon^*(\mathbf{B}, W, \delta)$ in $W$ as proven in Proposition 1 is evident. The windowed thresholds $\varepsilon^*(\mathbf{B}, W, \delta)$ for $\mathcal{C}_2$ and $\mathcal{C}_3$ are fairly close to the maximum windowed threshold even when $W = W_{min}$. The windowed thresholds for ensembles $\mathcal{C}_2$ and $\mathcal{C}_3$ are robust to changes in $\delta$, i.e. the thresholds are almost the same (the points overlap in the figure) for $\delta = 10^{-6}$ and $\delta = 10^{-12}$. Further, the windowed thresholds $\varepsilon^*(\mathcal{C}_i, W, \delta)$ are fairly close to the BP thresholds $\varepsilon^*_{BP}(\mathcal{C}_i), i = 1, 2, 3$ for $W \geq 12$. We will see next that this last observation is not always true. □

*Effect of termination:* The better BP performance of the $\mathcal{C}_c(J, K)$ ensemble in comparison with that of the $(J, K)$-regular block code ensemble (cf. Example 3) is because of the termination of the parity-check matrix of $\mathcal{C}_c(J, K)$ codes. More precisely, the low-degree CNs at the terminated portion of the protograph are more robust to erasures and their erasure-correcting power is cascaded through the rest of the protograph to give a better threshold for the convolutional ensemble in comparison with that for the corresponding unstructured ensemble [16]. From the definition of the WD, we can see that the sub-protograph within a window does not have the lower-degree checks if previous targeted symbols are not decoded. Therefore, we would expect a deterioration in the performance. Furthermore, the Design Rule 1 increases the degrees of the CNs in the terminated portion. Therefore, the effect of different termination on the WD performance is of interest.

*Example 6:* Tables I and II illustrate the WD thresholds for $\mathcal{C}_m(J, 2J)$ ensembles that satisfy Design Rule 1 except when $J = m_s + 1$. These ensembles are defined by the polynomials

$$p_1(x) = p_2(x) = (J - m_s) + x + x^2 + \cdots + x^{m_s}.$$

Note that $J \geq m_s + 1$. The ensembles are terminated so that the rate is $R_L = 0.49$. The worst threshold with WD (corresponding to the least window size $W_{min} = m_s + 1$) is denoted $\varepsilon^*_{m_s+1}$. The largest threshold with WD is denoted $\varepsilon^*_{L-m_s}$ and the BP threshold as $\varepsilon^*_{BP}$. The increase in the gap between $\varepsilon^*_{L-m_s}$ and $\varepsilon^*_{BP}$ with increasing $J$ illustrates the loss due to edge multiplicities (“weaker” termination). This is because the terminations at the beginning and at the end of the code are different, i.e. the CN degrees in the terminated portion at the beginning of the code are $2(J - m_s)$ which increases with $J$; whereas those at the end of the code are 2, a constant. Thus, much of the code performance is determined by the “stronger” (smaller check-degree) termination, the one at the end of the code for $J > m_s + 1$. This is also seen by the fact that the gap between $\varepsilon^*_{m_s+1}$ and $\varepsilon^*_{L-m_s}$ decreases as $J$ increases, meaning that the termination at the beginning of the code is weak and increasing the window size helps little. Note that the $\mathcal{C}_m(3, 6)$ ensemble in Table II is in fact the $\mathcal{C}_c(3, 6)$ classical ensemble, and that $\varepsilon^*_{L-m_s}$ is larger than the corresponding BP threshold. This is possible since WD only demands that the erasure probability of the targeted symbols is reduced to $\delta$. In contrast, BP demands that the erasure probability of all the symbols is reduced to 0. □

TABLE I
$m_s = 1, R_L = 0.49, \mathcal{C}_m(J, 2J), \delta = 10^{-12}$

| $J$ | $\varepsilon^*_{m_s+1}$ | $\varepsilon^*_{L-m_s}$ | $\varepsilon^*_{BP}$ |
|---|---|---|---|
| 2 | 0.0008 | 0.3162 | 0.3342 |
| 3 | 0.4499 | 0.4857 | 0.4872 |
| 4 | 0.4449 | 0.4469 | 0.4961 |
| 5 | 0.3915 | 0.3923 | 0.4969 |
| 6 | 0.3469 | 0.3475 | 0.4959 |
| 7 | 0.3115 | 0.3118 | 0.4891 |
| 8 | 0.2829 | 0.2832 | 0.4785 |
| 9 | 0.2595 | 0.2597 | 0.4666 |

TABLE II
$m_s = 2, R_L = 0.49, \mathcal{C}_m(J, 2J), \delta = 10^{-12}$

| $J$ | $\varepsilon^*_{m_s+1}$ | $\varepsilon^*_{L-m_s}$ | $\varepsilon^*_{BP}$ |
|---|---|---|---|
| 3 | 0.0189 | 0.4882 | 0.4876 |
| 4 | 0.4875 | 0.4947 | 0.4958 |
| 5 | 0.4493 | 0.4501 | 0.4971 |
| 6 | 0.3941 | 0.3945 | 0.4972 |
| 7 | 0.3489 | 0.3492 | 0.4969 |
| 8 | 0.3131 | 0.3133 | 0.4967 |
| 9 | 0.2843 | 0.2845 | 0.4957 |
| 10 | 0.2607 | 0.2608 | 0.4937 |

From the above discussion, we can add the following as another design rule.

*Design Rule 2:* For $\mathcal{C}_m(J, K)$ ensembles, keep the termination at the beginning of the code strong, preferably stronger than the one at the end of the code. That is, use polynomials $\mathcal{P} = \{p_j(x), j \in [K']\}$ such that each of the sums

$$\sum_{j=1}^{K'} p_j^{(0)}, \cdots, \sum_{j=1}^{K'} p_j^{(J'-1)}$$

is kept as small as possible.

*Targeted symbols:* We have thus far considered only the first $K'$ VNs in the sub-protograph contained within the window to be the targeted symbols. However, as an alternative way to trade-off performance for reduced latency, it is possible to consider other VNs also as targeted symbols. In this case, the window would be shifted beyond all the targeted symbols after processing each window configuration. For a window of size $W$, let us denote by $\varepsilon^*_i(\mathbf{B}, W, \delta)$ the windowed threshold when the targeted symbols are the first $iK'$ VNs, $1 \leq i \leq W$. Hence, $\varepsilon^*(\mathbf{B}, W, \delta) = \varepsilon^*_1(\mathbf{B}, W, \delta)$. By definition, $\varepsilon^*_i(\mathbf{B}, W, \delta) \leq \varepsilon^*(\mathbf{B}, W, \delta)$.

*Example 7:* Consider the $\mathcal{C}_m(6, 12)$ ensemble with $m_s = 1$ defined by $p_1(x) = p_2(x) = 3 + 3x$, denoted $\mathcal{C}_4$; and the $\mathcal{C}_m(4, 8)$ ensemble with $m_s = 1$ defined by $q_1(x) = q_2(x) = 2 + 2x$, denoted $\mathcal{C}_5$. Also consider ensembles $\mathcal{C}_6$ and $\mathcal{C}_7$ given by $r_1(x) = r_2(x) = 2 + 4x$ and $s_1(x) = s_2(x) = 2 + 2x + 2x^2$ respectively. Both $\mathcal{C}_6$ and $\mathcal{C}_7$ are $\mathcal{C}_m(6, 12)$ ensembles, but with memory $m_s = 1$ and 2 respectively. Table III gives the windowed thresholds $\varepsilon^*_i(\mathcal{C}_j, W = 4, \delta)$ with $iK'$ targeted symbols for a window of size 4 for $j = 4, 5, 6, 7$.

TABLE III
WINDOWED THRESHOLDS $\varepsilon^*_i(\mathcal{C}_j, W = 4, \delta = 10^{-12}), j = 4, 5, 6, 7$

| $i$ | $\mathcal{C}_4$ | $\mathcal{C}_5$ | $\mathcal{C}_6$ | $\mathcal{C}_7$ |
|---|---|---|---|---|
| 1 | 0.4429 | 0.4912 | 0.4835 | 0.4924 |
| 2 | 0.4429 | 0.4905 | 0.4835 | 0.4919 |
| 3 | 0.4427 | 0.4824 | 0.4828 | 0.4824 |
| 4 | 0.4294 | 0.3331 | 0.3331 | 0.3331 |

One might expect the windowed threshold $\varepsilon^*(\mathbf{B}, W, \delta)$ to be higher for an ensemble for which $\varepsilon^*_W(\mathbf{B}, W, \delta)$ is higher. This is not quite right: $\varepsilon_4^*(\mathcal{C}_4, 4, 10^{-12}) \approx 0.4294 > 0.3331 \approx \varepsilon_4^*(\mathcal{C}_5, 4, 10^{-12})$ whereas $\varepsilon_i^*(\mathcal{C}_5, 4, 10^{-12}) > \varepsilon_i^*(\mathcal{C}_4, 4, 10^{-12})\ \forall\ i < 4$. This can again be explained as the effect of stronger termination in $\mathcal{C}_5$ in comparison with $\mathcal{C}_4$. This is also evident in the larger thresholds for the $(6, 12)$ ensemble $\mathcal{C}_6$ with same memory as $\mathcal{C}_4$, but stronger termination. Also, keeping the same termination and increasing the memory improves the performance, as is exemplified by the larger thresholds of $\mathcal{C}_7$ in comparison with those of $\mathcal{C}_5$. □

The windowed thresholds $\varepsilon_i^*(\mathbf{B}, W, \delta)$ quantify the unequal erasure protection of different VNs in the sub-protograph within the window. Furthermore, it is clear that for good performance, it is advantageous to keep fewer targeted symbols within a window.

### *B. Finite length performance evaluation*

The finite length performance of LDPC codes under iterative message-passing decoding over the BEC is dependent on the number and the size of *stopping sets* present in the parity-check matrix of the code [23], [25]. Thus, the performance of the codes varies based on the parity-check matrix used to represent the code and, consequently, the performance of iterative decoding can be made to approach that of ML decoding by adding redundant rows to the parity-check matrix (See e.g. [26]). However, since we are exploiting the structure of the parity-check matrix of the convolutional code, we will not be interested in changing the parity-check matrix by adding redundant rows as this destroys the convolutional structure. The *ensemble stopping set size distribution* for some protograph-based LDPC codes was evaluated in [27] where it was shown that a minimum stopping set size that grows linearly in blocklength is important for the good performance of codes with short blocklengths. This analysis is similar to the analysis of the minimum distance growth rate of LDPC-CC ensembles—see [28] and references therein. It is worthwhile to note that although the minimum stopping set size grows linearly for protograph codes expanded using random permutation matrices, the same is not true for codes expanded using circulant permutation matrices [29]. In the following we will evaluate the finite length performance of codes constructed from $\mathcal{C}_m(J, K)$ ensembles with BP and WD through Monte Carlo simulations. WD was considered with only the first $K'M$ symbols as the targeted symbols.

In Figs. 4 and 5, the symbol error rate (SER) and the codeword error rate (CER) performance are depicted for codes $C_1 \in \mathcal{C}_1$ and $C_2 \in \mathcal{C}_2$, where the ensembles $\mathcal{C}_1$ and $\mathcal{C}_2$ were defined in Example 5. The codes used were those constructed by Liva [12] by expanding the protographs using circulant matrices (and sums of circulant matrices) and techniques of progressive edge growth (PEG) [30] and approximate cycle extrinsic message degree (ACE) [31] to avoid small cycles in the Tanner graphs of the codes. The girth of both the codes $C_1$ and $C_2$ was 12. The parameters used for the construction were $L = 20$ and $M = 512$ so that the blocklength $n = LK'M = 20480$ and $R_L = 0.45$. The BP thresholds for ensembles $\mathcal{C}_1$ and $\mathcal{C}_2$ with $L = 20$ were 0.4883 and 0.4882 respectively. As is clear from Figs. 4 and 5, code $C_2$ outperforms code $C_1$ for small window sizes ($W = 3, 5$), confirming the effectiveness of the proposed design rules for windowed decoding. For larger window sizes ($W = 10$), there is no marked difference in the performance of the two codes. It was also observed that for small $M$ values ($< 128$), the performance of codes constructed through circulant permutation matrices was better than those constructed through random permutation matrices. This difference in performance diminished for larger $M$ values.

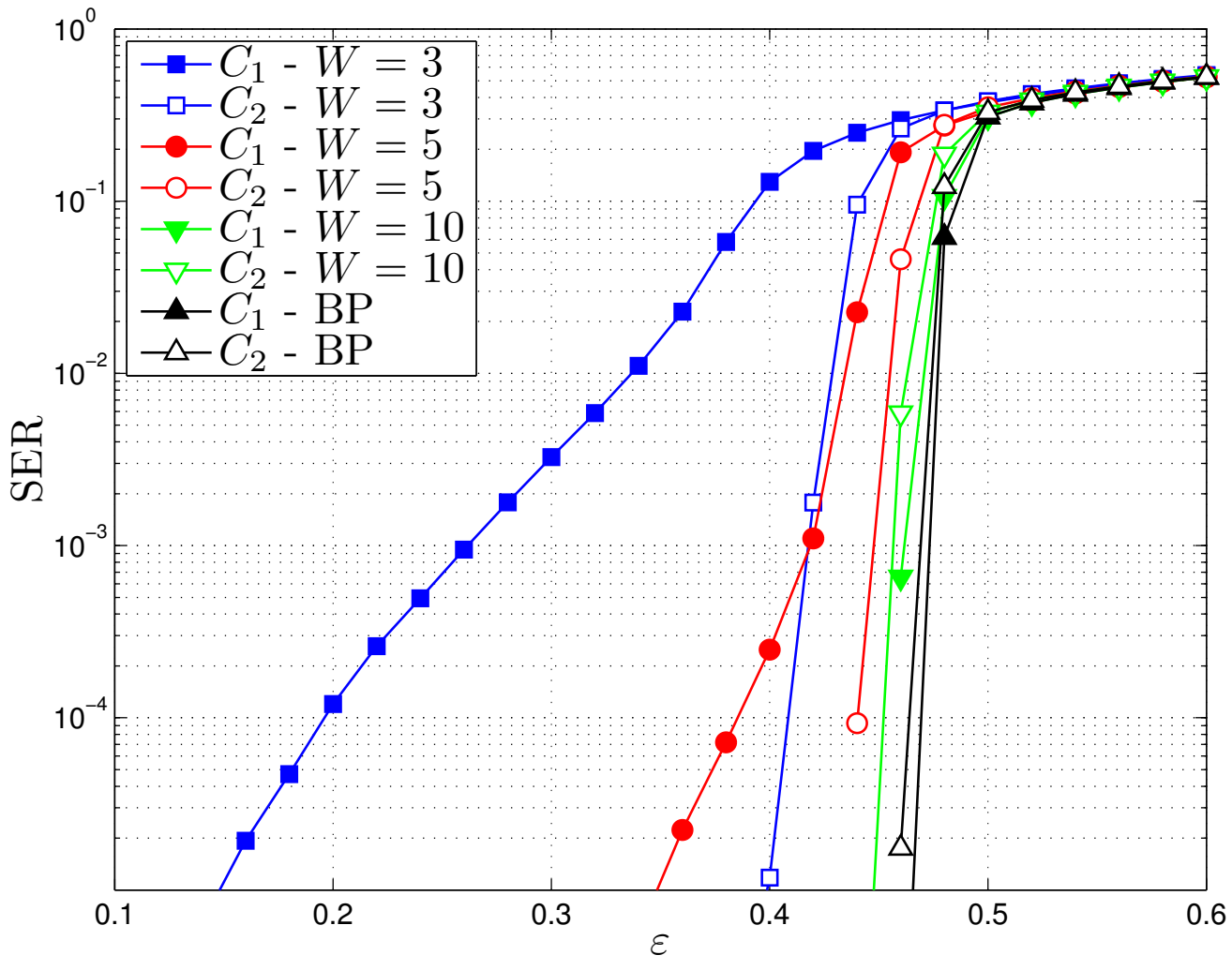

Fig. 4. SER performance for BP and Windowed Decoding over BEC.

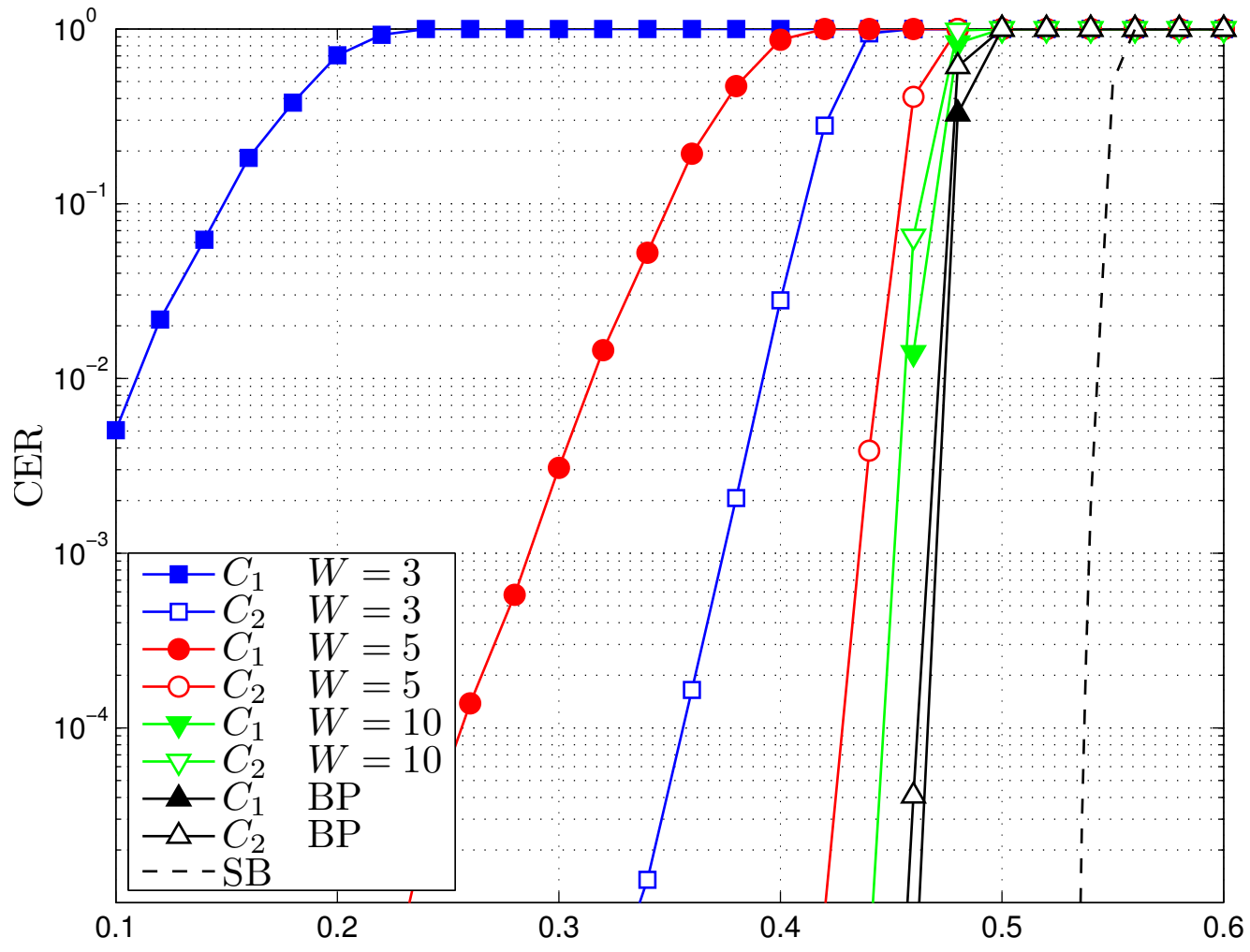

Fig. 5. CER performance for BP and Windowed Decoding over BEC. Also shown is the (Singleton) lower bound $\mathbb{P}_{SB}$ as SB.

We include in Fig. 5, for comparison, a lower bound on the CER $\mathbb{P}_{cw}$. The Singleton bound, $\mathbb{P}_{SB}$, represents the performance achievable by an idealized $(n, k)$ binary MDS code. This bound for the BEC can be expressed as

$$\mathbb{P}_{cw} \geq \sum_{j=n-k+1}^{n} \binom{n}{j} \varepsilon^j (1 - \varepsilon)^{n-j} = \mathbb{P}_{SB}.$$

Note that by the idealized $(n, k)$ binary MDS code, we mean a binary linear code that achieves the Singleton bound $d_{min} \leq$

$n-k+1$ with equality. This code does not exist for all values of $k$ and $n$.

## V. Erasure Channels with Memory

We now consider the performance of LDPC-CC ensembles and codes over erasure channels with memory. We consider the familiar two-state Gilbert-Elliott channel (GEC) [32], [33] as a model of an erasure channel with memory. In this model, the channel is either in a "good" state $G$, where we assume the erasure probability is 0, or in an "erasure" state $E$, in which the erasure probability is 1. The state process of the channel is a first-order Markov process with the transition probabilities $\mathbb{P}\{E \to G\} = g$ and $\mathbb{P}\{G \to E\} = b$. With these parameters, we can easily deduce [34] that the *average erasure rate* $\varepsilon$ and the *average burst length* $\Delta$ are given by

$$\varepsilon = \mathbb{P}\{E\} = \frac{b}{b+g}, \ \Delta = \frac{1}{g}.$$

We will consider the GEC to be parameterized by the pair $(\varepsilon, \Delta)$. Note that there is a one-to-one correspondence between the two pairs $(b, g)$ and $(\varepsilon, \Delta)$.

*Discussion:* The channel capacity of a correlated binary erasure channel with an average erasure rate of $\varepsilon$ is given as $(1-\varepsilon)$, which is the same as that of the memoryless channel, provided the channel is ergodic. Therefore, one can obtain good performance on a correlated erasure channel through the use of a capacity-achieving code for the memoryless channel with an interleaver to randomize the erasures [27], [35]. This is equivalent to permuting the columns of the parity-check matrix of the original code. We are not interested in this approach since such permutations destroy the convolutional structure of the code and as a result, we are unable to use the WD for such a scheme.

Construction of LDPC block codes for bursty erasure channels has been well studied. The performance metric of a code over a bursty erasure channel is related to the *maximum resolvable erasure burst length* (MBL) denoted $\Delta_{max}$ [35], which, as the name suggests, is the maximal length of a single solid erasure burst that can be decoded by a BP decoder. Methods of optimizing codes for such channels therefore focus on permuting columns of parity-check matrices to maximize $\Delta_{max}$, e.g. [36]–[41]. Instead of permuting columns of the parity-check matrix, in order to maintain the convolutional structure of the code, we will consider designing $\mathcal{C}_m(J, K)$ ensembles that maximize $\Delta_{max}$.

### A. Asymptotic Analysis

*1) **BP**:* As noted earlier, the performance of LDPC-CC ensembles depends on stopping sets. The structure of protographs imposes constraints on the code that limit the stopping set sizes and locations, as will be shown shortly.

Let us define a *protograph stopping set* to be a subset $S(\mathbf{B})$ of the VNs of the protograph $\mathbf{B}$ whose neighboring CNs are connected at least twice to $S(\mathbf{B})$. These are also denoted as $S(\mathcal{P})$, in terms of the set of polynomials defining the protograph. We define the *size* of the stopping set as the cardinality of $S(\mathbf{B})$, denoted $|S(\mathbf{B})|$. We call the least number of consecutive columns of $\mathbf{B}$ that contain the stopping set $S(\mathbf{B})$ the *span* of the stopping set, denoted $\langle S(\mathbf{B})\rangle$. Let us denote the size of the smallest protograph stopping set of the protograph $\mathbf{B}$ by $|S(\mathbf{B})|^*$, and the minimum number of consecutive columns of the protograph $\mathbf{B}$ that contain a protograph stopping set by $\langle S(\mathbf{B})\rangle^*$. When the protograph under consideration is clear from the context, we will drop it from the notation and use $|S|^*$ and $\langle S\rangle^*$. The minimum span of a stopping set is of interest because we can give simple bounds for $\Delta_{max}$ based on $\langle S(\mathbf{B})\rangle^*$. Note that the stopping set of minimal size and the stopping set of minimal span are not necessarily the same set of VNs. However, we always have

$$|S(\mathbf{B})|^* \leq \langle S(\mathbf{B})\rangle^*.$$

The following example clarifies the notation.

*Example 8:* Let us denote the base matrix corresponding to the protograph of the ensembles $\mathcal{C}_i$ of Example 5 as $\mathbf{B}^{(i)}, i = 1, 2, 3$. For ensembles $\mathcal{C}_1$ and $\mathcal{C}_3$, the first two columns of $\mathbf{B}^{(i)}, i = 1, 3$ form a protograph stopping set, i.e. $S(\mathbf{B}^{(i)}) = \{V_1, V_2\}, i = 1, 3$ is a stopping set. This is clear from the highlighted columns below

$$\mathbf{B}^{(1)} = \begin{pmatrix} \mathbf{1} & \mathbf{1} & 0 & 0 & 0 & 0 & \cdots \\ \mathbf{1} & \mathbf{1} & 1 & 1 & 0 & 0 & \cdots \\ \mathbf{1} & \mathbf{1} & 1 & 1 & 1 & 1 & \cdots \\ \mathbf{0} & \mathbf{0} & 1 & 1 & 1 & 1 & \cdots \\ \mathbf{0} & \mathbf{0} & 0 & 0 & 1 & 1 & \cdots \\ \vdots & \vdots & \vdots & \vdots & \vdots & \vdots & \ddots \end{pmatrix},$$

$$\mathbf{B}^{(3)} = \begin{pmatrix} \mathbf{2} & \mathbf{2} & 0 & 0 & 0 & 0 & \cdots \\ \mathbf{1} & \mathbf{1} & 2 & 2 & 0 & 0 & \cdots \\ \mathbf{0} & \mathbf{0} & 1 & 1 & 2 & 2 & \cdots \\ \mathbf{0} & \mathbf{0} & 0 & 0 & 1 & 1 & \cdots \\ \vdots & \vdots & \vdots & \vdots & \vdots & \vdots & \ddots \end{pmatrix}.$$

Therefore, $|S(\mathbf{B}^{(i)})|^* \leq 2$ and $\langle S(\mathbf{B}^{(i)})\rangle^* \leq 2$. Since no single column forms a protograph stopping set, $|S(\mathbf{B}^{(i)})|^* \geq 2$ and $\langle S(\mathbf{B}^{(i)})\rangle^* \geq 2$, implying $|S(\mathbf{B}^{(i)})|^* = \langle S(\mathbf{B}^{(i)})\rangle^* = 2, i = 1, 3$.

For ensemble $\mathcal{C}_2$, the highlighted columns of $\mathbf{B}^{(2)}$ in the following matrix form a protograph stopping set, i.e. $S(\mathbf{B}^{(2)}) = \{V_1, V_4\}$ is a stopping set.

$$\mathbf{B}^{(2)} = \begin{pmatrix} \mathbf{2} & 2 & 0 & \mathbf{0} & 0 & 0 & \cdots \\ \mathbf{0} & 1 & 2 & \mathbf{2} & 0 & 0 & \cdots \\ \mathbf{1} & 0 & 0 & \mathbf{1} & 2 & 2 & \cdots \\ \mathbf{0} & 0 & 1 & \mathbf{0} & 0 & 1 & \cdots \\ \mathbf{0} & 0 & 0 & \mathbf{0} & 1 & 0 & \cdots \\ \vdots & \vdots & \vdots & \vdots & \vdots & \vdots & \ddots \end{pmatrix}.$$

Thus, $|S(\mathbf{B}^{(2)})|^* \leq 2$ and $\langle S(\mathbf{B}^{(2)})\rangle^* \leq 4$. As no single column of $\mathbf{B}^{(2)}$ is a protograph stopping set and no three consecutive columns of $\mathbf{B}^{(2)}$ contain a protograph stopping set, it is clear that $|S(\mathbf{B}^{(2)})|^* \geq 2$ and $\langle S(\mathbf{B}^{(2)})\rangle^* \geq 4$, so that

$$2 = |S(\mathbf{B}^{(2)})|^* \leq \langle S(\mathbf{B}^{(2)})\rangle^* = 4.$$

$$\overline{\langle S_{l_1,l_2}\rangle} = \begin{cases} K'(m_s(l_1,l_2)-1)+(l_2-l_1+1), & i(l_1,l_2)=i_{l_2}\le i_{l_1}, j_{l_2}\le j_{l_1}=j(l_1,l_2) \\ K'(m_s(l_1,l_2)-1)+1, & i(l_1,l_2)=i_{l_2}\le i_{l_1}, j_{l_1}< j_{l_2}=j(l_1,l_2) \\ K'(m_s(l_1,l_2)-1)+1, & i(l_1,l_2)=i_{l_1}< i_{l_2}, j_{l_2}\le j_{l_1}=j(l_1,l_2) \\ K'(m_s(l_1,l_2)-1)-(l_2-l_1-1), & i(l_1,l_2)=i_{l_1}< i_{l_2}, j_{l_1}< j_{l_2}=j(l_1,l_2). \end{cases} \tag{7}$$

In these cases, it so happened that the stopping set with the minimal size and the stopping set with the minimal span were the same. □

Our aim in the following will be to obtain bounds for the maximal $\langle S(\mathbf{B})\rangle^*$ over $\mathcal{C}_m(J,K)$ ensembles with memory $m_s$, which we denote $\langle S(J,K,m_s)\rangle^*$, and design protographs that achieve minimal spans close to this optimal value.

The analysis of the minimal span of stopping sets for unstructured LDPC ensembles was performed in [42]. However, the structure of the protograph-based LDPC-CC allows us to obtain $\langle S(J,K,m_s)\rangle^*$ much more easily for some $\mathcal{C}_m(J,K)$ ensembles.

We start by observing that if one of the VNs in the protograph is connected multiple times to all its neighboring CNs, then it forms a protograph stopping set by itself. In order to obtain a larger minimum span of stopping sets, it is desirable to avoid this case, and we include this as one of our design criteria.

> *Design Rule 3:* For a $\mathcal{C}_m(J,K)$ ensemble, choose the polynomials $p_j(x)$ such that for every $j\in[K']$, there exists $0\le i_j\le(m_s+1)J'-1$ such that $p_j^{(i_j)}=1$.

Using the polynomial representation of LDPC-CC ensembles is helpful in this case since we can easily track stopping sets as those subsets that have polynomials whose coefficients are all larger than 1. From this fact, we can prove the following.

*Proposition 2 ($\langle S\rangle^*$ for $\mathcal{C}_m(J,2J)$ protographs):* For $\mathcal{C}_m(J,2J)$ protographs of memory $m_s$ defined by polynomials $p_1(x)$ and $p_2(x)$, $\langle S\rangle^*$ can be upper bounded as

$$\langle S\rangle^* \le \begin{cases} 2m_s, & 0=i_2\le i_1, j_2\le j_1=m_s \\ 2m_s-1, & 0=i_2\le i_1, j_1< j_2=m_s \\ 2m_s-1, & 0=i_1< i_2, j_2\le j_1=m_s \\ 2m_s-2, & 0=i_1< i_2, j_1< j_2=m_s \end{cases}$$

where $i_l=\min\deg(p_l(x))$ and $j_l=\deg(p_l(x)), l=1,2$. ■

We give the proof in Appendix II. We see from the above that $\langle S(J,2J,m_s)\rangle^*\le 2m_s$ and a necessary condition for achieving this span is the first of four possible cases listed above, which we include as another design criterion.

> *Design Rule 4:* For $\mathcal{C}_m(J,2J)$ ensembles with memory $m_s$, set
>
> $$\min\deg(p_2(x))=0 \text{ and } \deg(p_1(x))=m_s.$$

*Corollary 3 (Optimal $\mathcal{C}_m(J,2J)$ protographs):* For $\mathcal{C}_m(J,2J)$ protographs with memory $m_s$ and $J>2$, $\langle S(J,2J,m_s)\rangle^*=2m_s$. ■

The proof is given in Appendix III. Note that ensemble $\mathcal{C}_2$ in Example 5 achieves $\langle S(J,2J,m_s)\rangle^*$, as was observed in Example 8. It also satisfies design rules 1, 3 and 4. We bring to the reader's attention here that constructions other than the one given in the proof of the above corollary that achieve $\langle S\rangle^*=2m_s$ are also possible. These constructions allow us to design $\mathcal{C}_m(J,2J)$ ensembles for a wide range of required $\langle S\rangle^*$. We quickly see that a drawback of the convolutional structure is that if $m_s$ is increased to obtain a larger $\langle S\rangle^*$, the code rate $R_L$ decreases linearly for a fixed $L$.

We give without proof the following upper bound for $\langle S\rangle^*$ for $\mathcal{C}_m(J,K'J)$ ensembles, as it follows from Proposition 2.

*Proposition 4 ($\langle S\rangle^*$ for $\mathcal{C}_m(J,K'J)$ protographs):* For $\mathcal{C}_m(J,K'J)$ protographs defined by polynomials $\mathcal{P}=\{p_j(x), j\in[K']\}$, we have

$$\langle S\rangle^* \le \min_{(l_1,l_2)\in[K']^2, l_1<l_2}\{\overline{\langle S_{l_1,l_2}\rangle}\}$$

where $\overline{\langle S_{l_1,l_2}\rangle}$ is the upper bound for the minimal span $\langle S_{l_1,l_2}\rangle$ of stopping sets $S_{l_1,l_2}$ confined within subsets of the form $r_{l_1,l_2}(x)=a_1(x)p_{l_1}(x)+a_2(x)p_{l_2}(x)$ given in Equation (7), where we have used the notation $i_{l_u}=\min\deg(p_{l_u}(x))$, $j_{l_u}=\deg(p_{l_u}(x))$, $u=1,2$, $i(l_1,l_2)=\min\{i_{l_1},i_{l_2}\}$, $j(l_1,l_2)=\max\{j_{l_1},j_{l_2}\}$ and $m_s(l_1,l_2)=j(l_1,l_2)-i(l_1,l_2)$. ■

*Discussion:* By looking at the stopping sets confined within columns corresponding to two polynomials only, we can use Proposition 2 to upper bound the span of these stopping sets. The minimal such span over all possible choices of the two columns therefore gives an upper bound on the minimal span of the $(J,K'J)$ protograph. Since $\overline{\langle S_{l_1,l_2}\rangle}\le K'm_s\ \forall\, l_1,l_2$ from Equation (7), we have $\langle S(J,K'J,m_s)\rangle^*\le K'm_s$, which is similar to the result in Proposition 2. This bound is, however, loose in general.

For terminated codes, we can give an upper bound for $\langle S\rangle^*$ that is tighter in some cases.

*Corollary 5 ($\langle S\rangle^*$ for $\mathcal{C}_m(J,K)$ protographs):* For $\mathcal{C}_m(J,K)$ protographs terminated after $L$ instants, $\langle S\rangle^*\le K'L$.

*Proof:* From the Singleton bound for the protograph, we have $\langle S\rangle^*\le J'(L+m_s)$. Since we need $m_s\le\frac{R}{1-R}L$ for a positive code rate in (2), $\langle S\rangle^*\le\frac{J'L}{1-R}=K'L$.

Note that for $\mathcal{C}_m(J,K'J)$ protographs, this is tighter than the bound $\langle S\rangle^*\le K'm_s\le K'(K'-1)L$, which, in the worst case, is a factor $(K'-1)$ times larger. However, since we are interested mainly in ensembles for which $m_s\ll L$, this bound might be looser than the one in Proposition 4 for $\mathcal{C}_m(J,K'J)$ ensembles. ■

*Example 9:* Consider the $\mathcal{C}_m(J,K'J)$ ensemble with memory $m_s = u(K'-1)+1, m_s \leq (K'-1)L$ defined by the polynomials

$$p_l(x) = (J-1) + x^{j_l}, l \in [K'],$$

$j_l = m_s - u(l-1)$. It can be shown by an argument similar to the one used to prove Corollary 3 that for the protograph of this ensemble, $\langle S \rangle^* = K'u + 2$. This is exactly the bound in Proposition 4 since

$$\min_{l_1<l_2}\{\overline{\langle S_{l_1,l_2}\rangle}\} = \langle S_{K'-1,K'}\rangle = K'u + 2.$$

Thus, in this case

$$\langle S \rangle^* = \frac{K'}{K'-1}(m_s - 1) + 2 = \left\lceil \frac{K'}{K'-1} m_s \right\rceil$$

which is roughly only a fraction of the (loose) upper bound for $\langle S(J,K'J,m_s)\rangle^*$ suggested in the discussion of Proposition 4. The constructed $\mathcal{C}_m(J,K'J)$ protographs are thus optimal in the sense of maximizing the minimal span of stopping sets, i.e.

$$\langle S(J,K'J,u(K'-1)+1)\rangle^* = K'u + 2 \ \forall\ u \in [L-1].$$

They also satisfy Design Rules 1 and 3 for $J > 2$. Although Proposition 4 gave a tight bound for $\langle S \rangle^*$ in this case, it is loose in general. □

We can show that the $\mathcal{C}_m(J,K)$ protographs have minimal spans at least as large as the corresponding spans of $\mathcal{C}_m(a,K)$ protographs.

*Proposition 6:* $\langle S(J,K,m_s)\rangle^* \geq \langle S(a,K,m_s)\rangle^*$ where $a = \gcd(J,K) \geq 2$.

*Proof:* The equality is trivial when $a = J$. When $2 \leq a < J = aJ'$, one way of constructing the $\mathcal{C}_m(J,K)$ ensembles with memory $m_s$ is to let each set of modulo polynomials $\mathcal{P}_l$ themselves define $\mathcal{C}_m(a,K)$ ensembles with memory $m_s$. The result then follows by noting that a stopping set for the polynomials $\mathcal{P}$ has to be a stopping set for every set of polynomials $\mathcal{P}_l$, $l = 0, 1, \cdots, J'-1$. ■

The construction proposed above often allows us to strictly increase the minimal span of the $\mathcal{C}_m(J,K)$ ensemble in comparison with the $\mathcal{C}_m(a,K)$ ensemble, as illustrated by the following example.

*Example 10:* Consider the construction of a $\mathcal{C}_m(4,6)$ ensemble with memory 3. Let us call it $\mathcal{C}_8$. The different parameters in this case are $J = 4$, $K = 6$, $a = 2$, $J' = 2$, $K' = 3$ and $m_s = 3$. Since $m_s = u(K'-1)+1$ with $u = 1$, we have for $\mathcal{C}_m(2,6)$ protographs, $\langle S(2,6,3)\rangle^* = 5$ from Example 9 and we will define the modulo polynomials $\mathcal{P}_0$ to be the optimal construction that achieves this minimal span, i.e. $\mathcal{P}_0 = \{1+x^3, 1+x^2, 1+x\}$. Then, by defining $\mathcal{P}_1 = \{1+x^3, 1+x^3, 1+x^3\}$, we can show that $\langle S(\mathcal{P})\rangle^* = 6$ and hence

$$\langle S(4,6,3)\rangle^* \geq 6 > 5 = \langle S(2,6,3)\rangle^*.$$

Note that the protograph defined by $\mathcal{P}$ has no degree-1 VNs associated with the component matrix $\mathbf{B}_0$. In fact, the constructed $\mathcal{C}_m(4,6)$ ensemble has $\varepsilon^*(\mathcal{C}_8, m_s+1, 10^{-12}) \approx 0.6469$, fairly close to the Shannon limit of $\varepsilon^{Sh} = \frac{2}{3}$, even with the smallest possible window size. Table IV lists the windowed thresholds of this ensemble with different numbers of targeted symbols within the smallest window for $\delta = 10^{-12}$. □

TABLE IV
$\varepsilon_i^*(\mathcal{C}_8, m_s+1, 10^{-12})$

| $i$ | 1 | 2 | 3 | 4 |
|---|---|---|---|---|
| $\varepsilon_i^*$ | 0.6469 | 0.6184 | 0.5803 | 0.4997 |

*2) **WD**:* The asymptotic analysis for WD is essentially the same as that for BP. We will consider WD with only the first $K'$ symbols within each window as the targeted symbols. We are now interested in the sub-protograph stopping sets, denoted $S(\mathbf{B}, W)$, that include one or more of the targeted symbols within a window. Let us denote the minimal span of such stopping sets as $\langle S(\mathbf{B}, W)\rangle^*$. Since stopping sets of the protograph of the LDPC-CC are also stopping sets of the sub-protograph within a window, and since such stopping sets can be chosen to include some targeted symbols within the window, we have $\langle S(\mathbf{B}, W)\rangle^* \leq \langle S(\mathbf{B})\rangle^*$. In fact, $\langle S(\mathbf{B}, W)\rangle^* = \langle S(\mathbf{B})\rangle^*$ when

$$W \geq \left\lceil \frac{\langle S \rangle^*}{K'} \right\rceil + m_s$$

since in this case the first $K' \left\lceil \frac{\langle S \rangle^*}{K'} \right\rceil$ columns are completely contained in the window. Further, we have

$$\langle S(\mathbf{B}, W)\rangle^* \leq \langle S(\mathbf{B}, W+1)\rangle^*.$$

This is true because a stopping set for window size $W$ involving targeted symbols is not necessarily a stopping set for window size $W+1$, whereas a stopping set for window size $W+1$ is definitely a stopping set for window size $W$.

*Remark:* When the first $iK'$ symbols within a window are the targeted symbols, we have for $i \leq W - m_s$

$$\langle S_i(\mathbf{B}, W)\rangle^* = \langle S(\mathbf{B}, W-i+1)\rangle^*$$

where $\langle S_i(\mathbf{B}, W)\rangle^*$ denotes the minimal span of stopping sets of the sub-protograph within the window of size $W$ involving at least one of the $iK'$ targeted symbols, and $\langle S_1(\mathbf{B}, W)\rangle^* = \langle S(\mathbf{B}, W)\rangle^*$. Consequently, we have

$$\langle S_i(\mathbf{B}, W)\rangle^* \leq \langle S(\mathbf{B}, W)\rangle^*.$$

The definition of $\langle S_i(\mathbf{B}, W)\rangle^*$ can be extended to accommodate $W - m_s + 1 \leq i \leq W$, as in the case of windowed thresholds. In particular, we have $\langle S_W(\mathbf{B}, W)\rangle^* = \langle S(\mathbf{B}_0)\rangle^* \leq J'$, where the last inequality is from the Singleton bound.

*Example 11:* Consider the ensemble $\mathcal{C}_2$ defined in Example 5. With a window of size $W = m_s + 1 = 3$, we have $\langle S(\mathcal{C}_2, 3)\rangle^* = 2$ with the corresponding stopping set $S_3 = \{V_2, V_3\}$ highlighted below

$$\begin{pmatrix} 2 & \mathbf{2} & \mathbf{0} & 0 & 0 & 0 \\ 0 & \mathbf{1} & \mathbf{2} & 2 & 0 & 0 \\ 1 & \mathbf{0} & \mathbf{0} & 1 & 2 & 2 \end{pmatrix}$$

and with a window size $W = 4$, we have $\langle S(\mathcal{C}_2, 4)\rangle^* = \langle S(\mathcal{C}_2)\rangle^* = 4$, and the corresponding stopping sets $S_4 = \{V_1, V_4\}$ and $S_4' = \{V_1, V_2, V_4\}$ are as follows

$$\begin{pmatrix} \mathbf{2} & 2 & 0 & \mathbf{0} & 0 & 0 & 0 & 0 \\ \mathbf{0} & 1 & 2 & \mathbf{2} & 0 & 0 & 0 & 0 \\ \mathbf{1} & 0 & 0 & \mathbf{1} & 2 & 2 & 0 & 0 \\ \mathbf{0} & 0 & 1 & \mathbf{0} & 0 & 1 & 2 & 2 \end{pmatrix},$$

$$\begin{pmatrix} \mathbf{2} & \mathbf{2} & 0 & \mathbf{0} & 0 & 0 & 0 & 0 \\ \mathbf{0} & \mathbf{1} & 2 & \mathbf{2} & 0 & 0 & 0 & 0 \\ \mathbf{1} & \mathbf{0} & 0 & \mathbf{1} & 2 & 2 & 0 & 0 \\ \mathbf{0} & \mathbf{0} & 1 & \mathbf{0} & 0 & 1 & 2 & 2 \end{pmatrix}.$$

Note that for window size 3, whereas the minimal span of a stopping set involving VN $V_2$ is 2, that of a stopping set involving $V_1$ is 4. However, for window size 4, the stopping set involving $V_1$ with minimal span, denoted $S_4$, and that involving $V_2$, $S_4'$, each have a span of 4, although their cardinalities are 2 and 3 respectively. We have in this case, $S_4 \subset S_4'$. Notice that $\langle S_2(\mathcal{C}_2, 4)\rangle^* = \langle S(\mathcal{C}_2, 3)\rangle^* = 2$. □

### *B. Finite length analysis*

*1) **BP**:* We now show the relation between the parameters $\Delta_{max}$ and $\langle S(\mathbf{B})\rangle^*$. We shall assume in the following that $\langle S\rangle^* \geq 2$, i.e. every column of the protograph has at least one of the entries equal to 1. We will consider the expansion of the protographs by a factor $M$ to obtain codes.

*Proposition 7:* For any $(J, K)$ regular LDPC-CC, $\Delta_{max} \leq M\langle S\rangle^* - 1$.

*Proof:* Clearly, the set of the $M\langle S\rangle^*$ columns of the parity-check matrix corresponding to the $\langle S\rangle^*$ consecutive columns of $\mathbf{B}$ that contain the protograph stopping set with minimal span must contain a stopping set of the parity-check matrix. Therefore, if all symbols corresponding to these columns are erased, they cannot be retrieved. ■

*Corollary 8:* A terminated $\mathcal{C}_m(J, K'J)$ LDPC-CC with $m_s = u(K'-1)+1, u \in [L-1]$ can never achieve the MBL of an MDS code.

*Proof:* From the Singleton bound, we have $\Delta_{max} \leq n - k = (L + m_s)M$, assuming that the parity-check matrix is full-rank. From Proposition 7 we have,

$$\begin{aligned}\Delta_{max} &\leq M\langle S(J, K'J, u(K'-1)+1)\rangle^* - 1 \\ &= \left\lceil \frac{K'}{K'-1} m_s \right\rceil M - 1\end{aligned}$$

where the second equality follows from the discussion in Example 9. Since we require $m_s \leq \frac{R}{1-R}L = (K'-1)L$ for a non-negative code rate in (2),

$$\begin{aligned}\Delta_{max} &\leq \left\lceil \frac{(K'-1)m_s + (K'-1)L}{K'-1} \right\rceil M - 1 \\ &< (L + m_s)M\end{aligned}$$

which shows that the MBL of an MDS code can never be achieved. ■

*Remark:* Although the idealized binary $(n, k)$ MDS code does not exist, there are codes that achieve MDS performance when used over a channel that introduces a single burst of erasures in a codeword. For example, the $(2n, n)$ code with a parity-check matrix $\mathbf{H} = [\mathbf{I}_n \ \mathbf{I}_n]$ has an MBL of $\Delta_{max} = n$.

Despite the discouraging result from Corollary 8, we can guarantee an MBL that linearly increases with $\langle S\rangle^*$ as follows.

*Proposition 9:* For any $(J, K)$ regular LDPC-CC, $\Delta_{max} \geq M(\langle S\rangle^* - 2) + 1$.

*Proof:* From the definition of $\langle S\rangle^*$, it is clear that if one of the two extreme columns is completely known, all other symbols can be recovered, for otherwise the remaining columns within the span of the stopping set $S$ will have to contain another protograph stopping set, violating the minimality of the stopping set span $\langle S\rangle^*$ (The two extreme columns are *pivots* of the stopping set [41].) The largest solid burst that is guaranteed to have at least one of the extreme columns completely known is of length $M(\langle S\rangle^* - 2) + 1$. Therefore, $\Delta_{max} \geq M(\langle S\rangle^* - 2) + 1$. ■

*Example 12:* For the $\mathcal{C}_m(J, K'J)$ ensemble with memory $m_s = u(K'-1)+1, u \in [L-1]$ in Example 9, we have

$$MK'\left(\frac{m_s - 1}{K'-1}\right) + 1 \leq \Delta_{max} \leq MK'\left(\frac{m_s - 1}{K'-1}\right) + 2M - 1$$

from Propositions 7 and 9. Thus, we can construct codes with MBL proportional to $m_s$. □

*2) **WD**:* The MBL for WD $\Delta_{max}(W)$ can be bounded as in the case of BP based on $\langle S(\mathbf{B}, W)\rangle^*$. Assuming that the window size is $W \geq m_s + 1$, the targeted symbols are the first $K'$ symbols within the window, and the polynomials defining the ensemble are chosen to satisfy Design Rule 3, we have $\langle S(\mathbf{B}, W)\rangle^* \geq 2$. Propositions 7 and 9 in this case imply that

$$M(\langle S(\mathbf{B}, W)\rangle^* - 2) + 1 \leq \Delta_{max}(W) \leq M\langle S(\mathbf{B}, W)\rangle^* - 1.$$

### *C. Numerical results*

The MBL for codes $C_1$ and $C_2$ (the same codes used in Section IV-B) was computed using an exhaustive search algorithm, by feeding the decoder with a solid burst of erasures and testing all the possible locations of the burst. The MBL for the codes we considered was 1023 and 1751 for codes $C_1$ and $C_2$, respectively. Note that for code $C_1$, the MBL $\Delta_{max} = 1023 = 2M - 1$, i.e., code $C_1$ achieves the upper bound from Proposition 7. More importantly, the maximum possible $\Delta_{max}$ was achievable while maintaining good performance over the BEC with the BP decoder. However, the MBL for code $C_2$, $\Delta_{max} = 1751 < 2047 = 4M - 1$, is much smaller than the corresponding bound from Proposition 7. In this case, although other code constructions with $\Delta_{max}$ up to 2045 were possible, a trade-off between the BEC performance and MBL was observed, i.e. the code that achieved $\Delta_{max} = 2045$ was found to be much worse over the BEC than both codes $C_1$ and $C_2$ considered here. Such a trade-off has also been observed by others, e.g. [40]. This could be because the codes that achieve large $\Delta_{max}$ are often those that have a very regular structure in their parity-check matrices. Nevertheless, our code design does give a large increase in MBL ($> 70\%$) when compared with the corresponding codes constructed from $\mathcal{C}_c$ ensembles, without any decrease in code rate (same $m_s$). The MBL achieved as a fraction of the maximum possible MBL $\Delta_{max}/(n-k)$ was roughly $9.1\%$ and $15.5\%$ for codes $C_1$ and $C_2$, respectively.

In Figs 6, 7 and 8, we show the CER performance obtained for codes $C_1$ and $C_2$ over GEC channels with $\Delta = 10, 50$ and 100 respectively, and $\varepsilon \in [0.1, 0.6]$. As can be seen from the figures, for $W = 3$, code $C_2$ always outperforms code $C_1$, while for $W = 5$ there is no such gain when $\Delta = 100$. However, for $W = 10$ and for BP decoding, code $C_1$ slightly outperforms $C_2$.

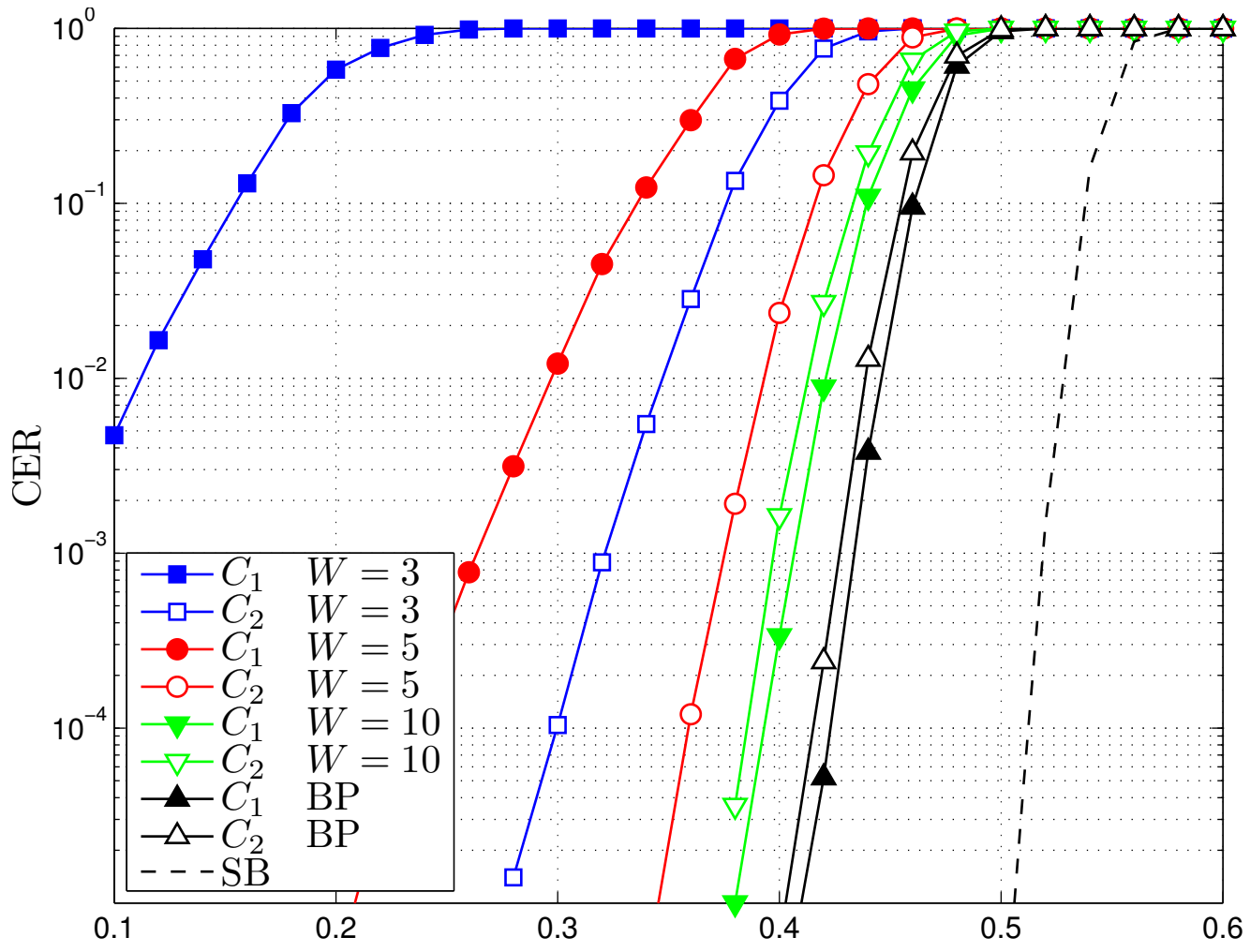

Fig. 6. CER Performance on GEC with $\Delta = 10$ with Singleton bound (SB).

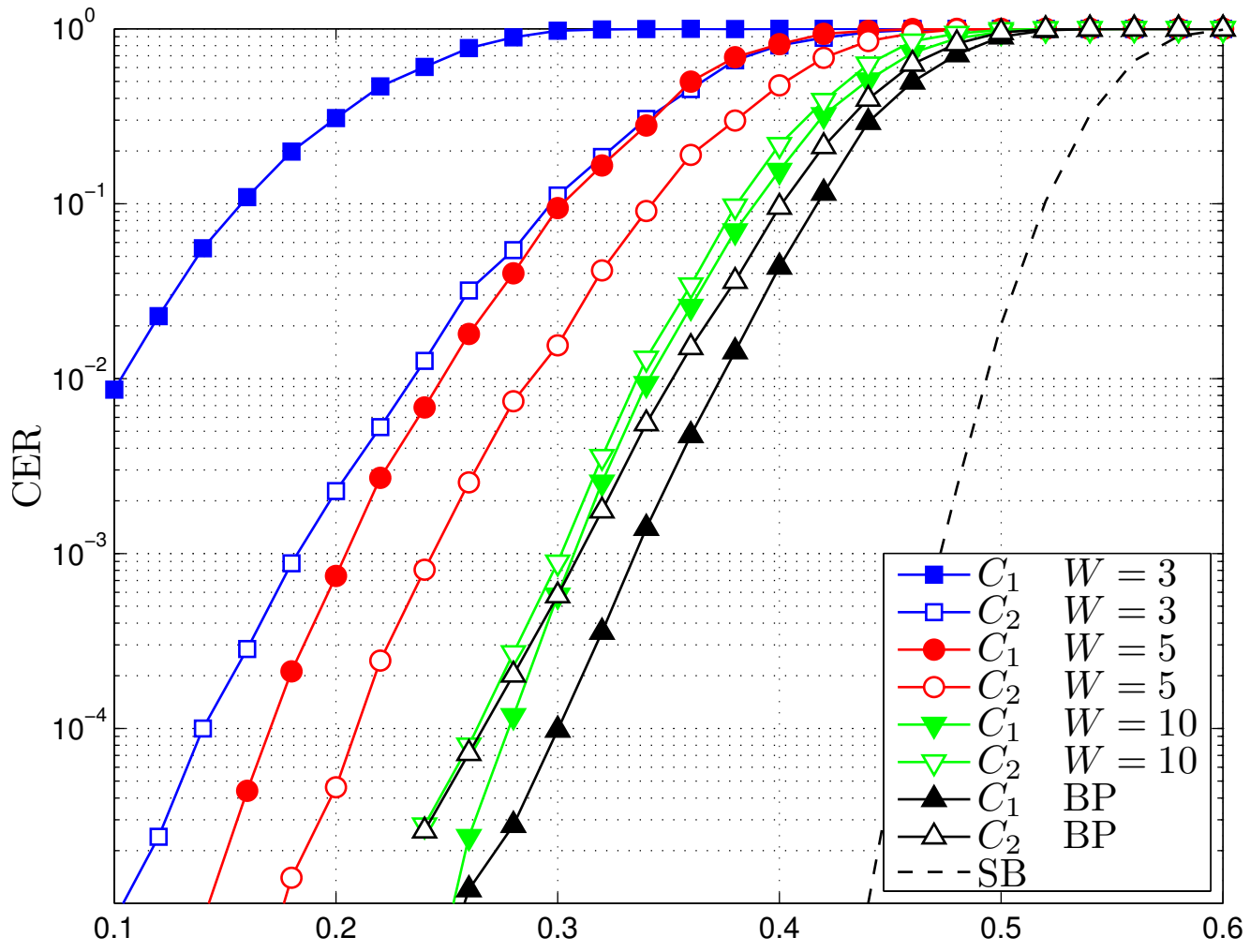

Fig. 7. CER Performance on GEC with $\Delta = 50$ with Singleton bound (SB).

Note that the code $C_2$ outperforms $C_1$ for small $\varepsilon$ when the average burst length $\Delta = 100$ for large window sizes and for BP decoding. This can be explained because in this regime, the probability of a burst is small but the average burst length is large. Therefore, when a burst occurs, it is likely to resemble a single burst in a codeword, and in this case we know that the code $C_2$ is stronger than $C_1$. Also note the significant gap between the BP decoder performance and the Singleton bound, suggesting that unlike some moderate length LDPC block codes with ML decoding [38], LDPC-CC are far from achieving MDS performance with BP or windowed decoding.

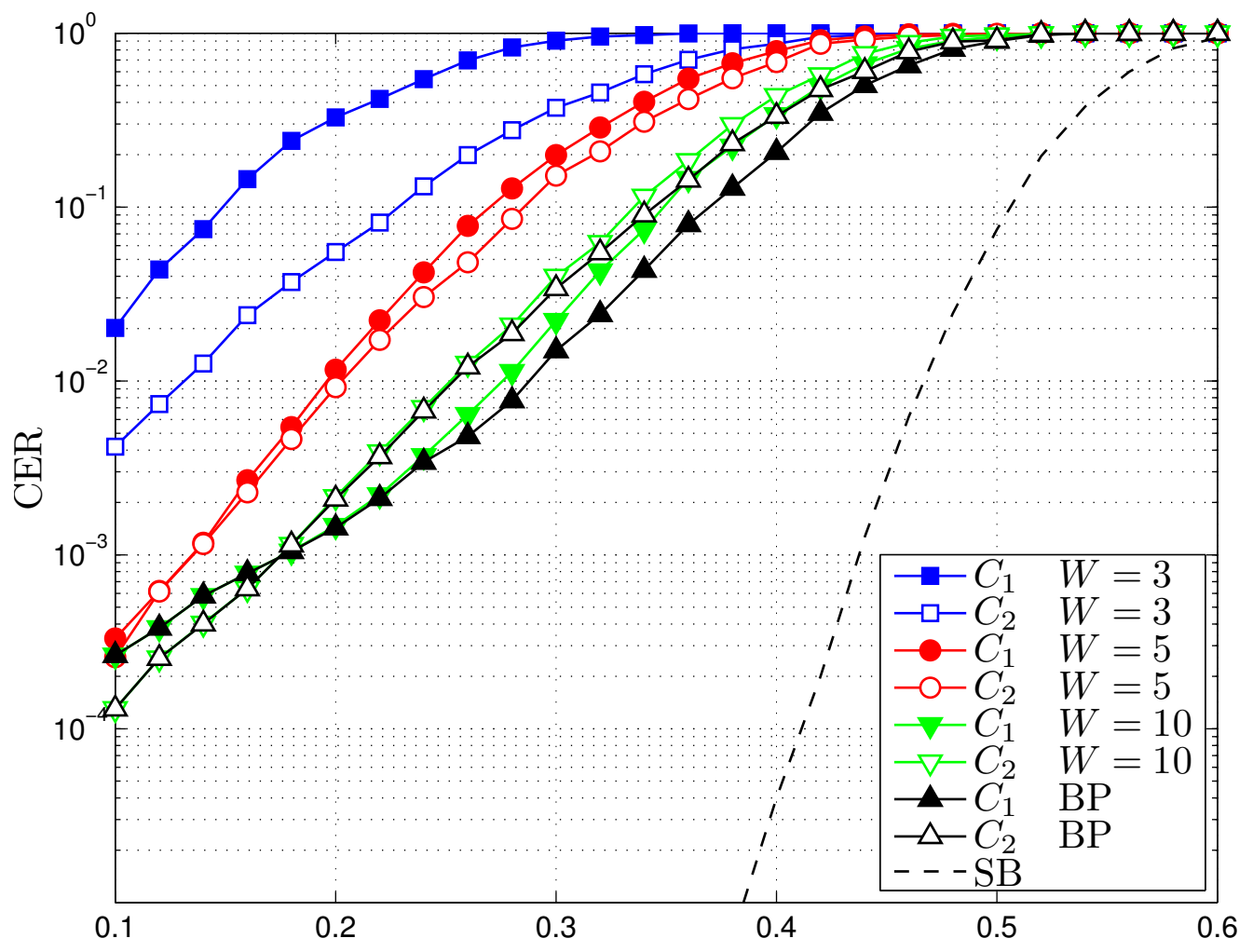

Fig. 8. CER Performance on GEC with $\Delta = 100$ with Singleton bound (SB).

## VI. Conclusions

We studied the performance of a windowed decoding scheme for LDPC convolutional codes over erasure channels. We showed that this scheme, when used to decode terminated LDPC-CC, provides an efficient way to trade-off decoding performance for reduced latency. Through asymptotic performance analysis, several design rules were suggested to avoid bad structures within protographs and, in turn, to ensure good thresholds. For erasure channels with memory, the asymptotic performance analysis led to design rules for protographs that ensure large stopping set spans. Examples of LDPC-CC ensembles that satisfy design rules for the BEC as well as erasure channels with memory were provided. Finite length codes belonging to the constructed ensembles were simulated and the validity of the design rules as markers of good performance was verified. The windowed decoding scheme can be used to decode LDPC-CC over other channels that introduce errors and erasures, although in this case error propagation due to wrong decoding within a window will have to be carefully dealt with.

For erasure channels, while close-to-optimal performance (in the sense of approaching capacity) was achievable for the BEC, we showed that the structure of LDPC-CC imposed constraints that bounded the performance over erasure channels with memory strictly away from the optimal performance (in the sense of approaching MDS performance). Nevertheless, the simple structure and good performance of these codes, as well as the latency flexibility and low complexity of the decoding algorithm, are attractive characteristics for practical systems.

## Acknowledgment

The authors are very grateful to the anonymous reviewers for their comments and suggestions for improving the presentation of the paper. They would also like to thank Gianluigi

Liva who suggested the trade-off of decoding performance for reduced latency through the use of a windowed decoder, and Rüdiger Urbanke for pointing out an error in an earlier version of the paper.

## Appendix I
## Proof of Proposition 1

Consider the $i^{\text{th}}$ window configuration for window sizes $W$ and $W+1$ shown in Fig. 9. We are interested in a window configuration that is not at the terminated portion of the code. Call the Tanner graphs of these windows $A = (V_A, C_A, E_A)$ and $B = (V_B, C_B, E_B)$ respectively, where $V_A, V_B$ and $C_A, C_B$ are the sets of VNs and CNs respectively and $E_A, E_B$ are the sets of edges. Clearly, $V_A \subset V_B, C_A \subset C_B$, and $E_A \subset E_B$. Any VN in $V_A$ that is connected to some variable in $V_B \setminus V_A$ has to be connected via some CN in $C_B \setminus C_A$. The edges between these CNs and VNs in $V_A$ are shown hatched in Fig. 9. Consider the computation trees for the a-posteriori message at a targeted symbol in $V_A$ and that for the same symbol in $V_B$. Call them $\mathcal{T}_A$ and $\mathcal{T}_B$ respectively. Then we have $\mathcal{T}_A \subset \mathcal{T}_B$.

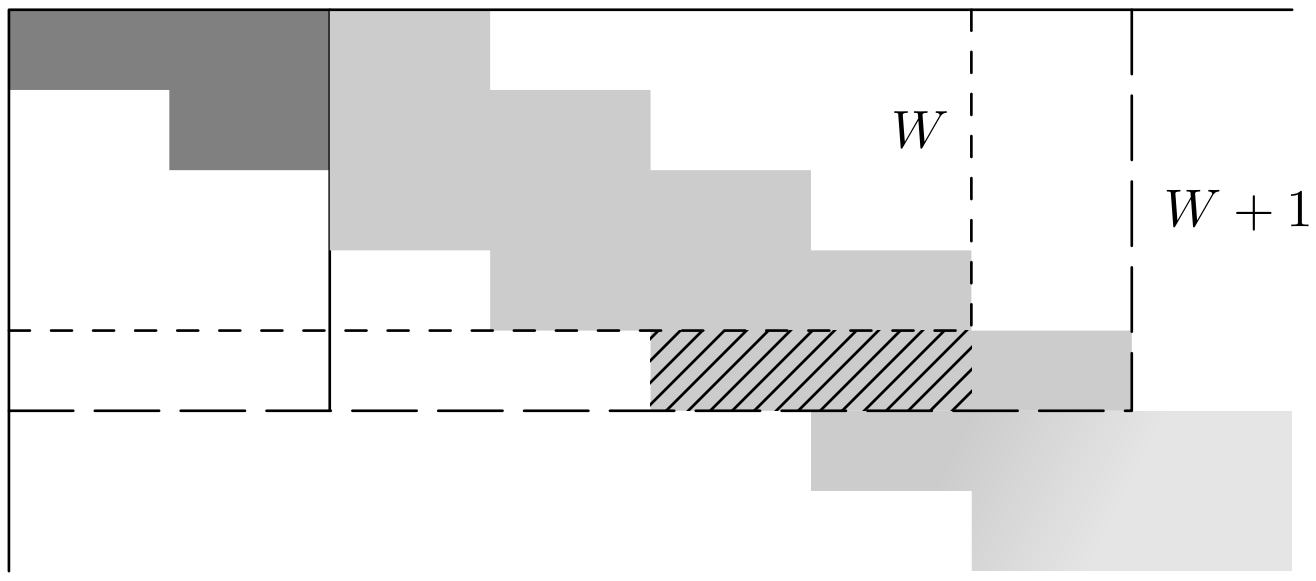

Fig. 9. Sub-protographs of window sizes $W$ and $W+1$. The edges connected to targeted symbols from previous window configurations are shown in darker shade of gray.

We now state two lemmas which will be made use of subsequently. The proofs of these lemmas are straightforward and have been omitted.

*Lemma 10 (Monotonicity of* $\mathtt{C}$*):* The CN operation in (5) is monotonic in its arguments, i.e.,

$$0 \leq x \leq x' \leq 1 \Rightarrow \mathtt{C}(x,y) \leq \mathtt{C}(x',y) \ \forall \ y \in [0,1],$$

where the two-argument function $\mathtt{C}(x,y) = xy$. ■

*Lemma 11 (Monotonicity of* $\mathtt{V}$*):* The VN operation in (6) is monotonic in its arguments, i.e.,

$$0 \leq x \leq x' \leq 1 \Rightarrow \mathtt{V}(x,y) \leq \mathtt{V}(x',y) \ \forall \ y \in [0,1],$$

where $\mathtt{V}(x,y) = 1-(1-x)(1-y)$. ■

The operational significance of the above lemmas is the following: if we can upper (lower) bound the mutual information on some incoming edge of a CN or a VN, and use the bound to compute the outgoing mutual information from that node, we get an upper (lower) bound on the actual outgoing mutual information. Thus, by bounding the mutual information on some edges of a computation tree and repetitively applying Lemmas 10 and 11, one can obtain bounds for the a-posteriori mutual information at the root of the tree.

We start by augmenting $\mathcal{T}_A$, creating another computation tree $\mathcal{T}_A^+$ that has the same structure as $\mathcal{T}_B$. In particular, $\mathcal{T}_A^+$ includes the additional edges corresponding to the hatched region. In $\mathcal{T}_A^+$ and $\mathcal{T}_B$, we denote the set of these edges by $E_u(\mathcal{T}_A^+)$ and $E_u(\mathcal{T}_B)$ respectively. In $\mathcal{T}_A^+$, we assign zero mutual information to each edge in $E_u(\mathcal{T}_A^+)$.

Now, let $I^{\mathcal{T}_A}$, $I^{\mathcal{T}_A^+}$ and $I^{\mathcal{T}_B}$ be the a-posteriori mutual information at the roots of the trees $\mathcal{T}_A$, $\mathcal{T}_A^+$ and $\mathcal{T}_B$ respectively. Then it is clear that $I^{\mathcal{T}_A} = I^{\mathcal{T}_A^+}$, since the messages on edges in $E_u(\mathcal{T}_A^+)$ are effectively erasures and zero out the contributions from the checks in $C_A^+ \setminus C_A = C_B \setminus C_A$.

On the other hand, if we denote by $I_e(\mathcal{T}_B)$ the mutual information associated with an edge $e \in E_u(\mathcal{T}_B)$, and by $I_e(\mathcal{T}_A^+)$ the mutual information associated with the corresponding edge in $\mathcal{T}_A^+$, we know that $I_e(\mathcal{T}_A^+) = 0$ so that $I_e(\mathcal{T}_A^+) \leq I_e(\mathcal{T}_B)$. Hence, we have from Lemmas 10 and 11 that $I^{\mathcal{T}_A^+} \leq I^{\mathcal{T}_B}$.

Since $I^{\mathcal{T}_A} = I^{\mathcal{T}_A^+}$, it follows that $I^{\mathcal{T}_A} \leq I^{\mathcal{T}_B}$, as desired.

## Appendix II
## Proof of Proposition 2

From the definitions made in the statement of Proposition 2, we have $0 \leq i_l < j_l \leq m_s, l = 1,2$. We assume $i_l < j_l$ in order to satisfy Design Rule 3. Since the code has memory $m_s$, we have $i = \min\{i_1, i_2\} = 0$ and $j = \max\{j_1, j_2\} = m_s$. Consider the subset of columns of $\mathbf{B}$ corresponding to the polynomial $r(x) = p_1(x)b_1(x) + p_2(x)b_2(x)$ where

$$b_1(x) = \begin{cases} x^{i_2}, & i_2 = j_2 - 1 \\ x^{i_2} + x^{i_2+1} + \cdots + x^{j_2-1}, & i_2 < j_2 - 1 \end{cases}$$

and

$$b_2(x) = \begin{cases} x^{i_1}, & i_1 = j_1 - 1 \\ x^{i_1} + x^{i_1+1} + \cdots + x^{j_1-1}, & i_1 < j_1 - 1. \end{cases}$$

We claim that this is a protograph stopping set. To see this, consider the columns corresponding to the above subset with $\beta(p_1(x))$ and $\beta(p_2(x))$ as the column polynomials defining $\mathbf{B}$. We have

$$\begin{aligned} \hat{r}(x) &= \beta(p_1(x))b_1(x) + \beta(p_2(x))b_2(x) \\ &= (x^{i_1} + x^{j_1})(x^{i_2} + x^{i_2+1} + \cdots + x^{j_2-1}) \\ &\quad + (x^{i_2} + x^{j_2})(x^{i_1} + x^{i_1+1} + \cdots + x^{j_1-1}) \\ &= x^{i_1+i_2} + x^{i_1+i_2+1} + \cdots + x^{i_1+j_2-1} \\ &\quad + x^{j_1+i_2} + x^{j_1+i_2+1} + \cdots + x^{j_1+j_2-1} \\ &\quad + x^{i_1+i_2} + x^{i_1+i_2+1} + \cdots + x^{j_1+i_2-1} \\ &\quad + x^{i_1+j_2} + x^{i_1+j_2+1} + \cdots + x^{j_1+j_2-1} \\ &= 2x^{i_1+i_2} + \cdots + 2x^{i_1+j_2-1} \\ &\quad + 2x^{i_1+j_2} + \cdots + 2x^{j_1+j_2-1} \end{aligned}$$

when $j_l > i_l + 1, l = 1,2$. Similarly, it can be verified that $\hat{r}(x)$ has all coefficients equal to 2 in all other cases also. Clearly, $\hat{r}(x) \preceq r(x)$ and thus $r(x)$ can only differ from $\hat{r}(x)$ in having larger coefficients. Therefore, $r(x)$ also has all coefficients greater than 1. This shows that the chosen subset of columns form a protograph stopping set. Based on the parameters $i_l, j_l, l = 1, 2$, we can count the number of columns included in the span of this stopping set and therefore give upper bounds on $\langle S \rangle^*$ as claimed :

$$\langle S \rangle^* \leq \begin{cases} 2(j_1 - i_2), & 0 = i_2 \leq i_1, j_2 \leq j_1 = m_s \\ 2(j_2 - i_2) - 1, & 0 = i_2 \leq i_1, j_1 < j_2 = m_s \\ 2(j_1 - i_1) - 1, & 0 = i_1 < i_2, j_2 \leq j_1 = m_s \\ 2(j_2 - i_1 - 1), & 0 = i_1 < i_2, j_1 < j_2 = m_s. \end{cases}$$

## Appendix III
## Proof of Corollary 3

Consider the protograph of the ensemble given by $p_1(x) = (J-1) + x^{m_s}$ and $p_2(x) = (J-1) + x$. Let the polynomial $r(x) = p_1(x)a_1(x) + p_2(x)a_2(x)$ represent an arbitrary subset (chosen from the $2^{2m_s-1} - 1$ non-empty subsets) of the first $(2m_s - 1)$ columns of $\mathbf{B}$, for any choice of polynomials $a_1(x)$ and $a_2(x)$ with coefficients in $\{0, 1\}$ and maximal degrees $(m_s - 1)$ and $(m_s - 2)$ respectively:

$$a_i(x) = \sum_{j=0}^{d_i} a_i^{(j)} x^j, i = 1, 2, \;\; d_1 = m_s - 1, d_2 = m_s - 2$$

where $a_i^{(j)} \in \{0, 1\}$ and not all $a_i^{(j)}$s are zeros. When $a_1(x) \neq 0$, let $i_1 = \deg(a_1(x))$. Clearly, $r(x)$ is a monic polynomial of degree $(m_s + i_1)$. When $a_1(x) = 0$ and $a_2(x) \neq 0$, let $i_2 = \deg(a_2(x))$. Then, $r(x)$ is a monic polynomial of degree $(1 + i_2)$. Since in both these cases $r(x)$ is a monic polynomial, there is at least one coefficient equaling 1. Thus, $\langle S \rangle^* > 2m_s - 1$. Finally, notice that

$$\begin{aligned} p_1(x) + x^{m_s-1}p_2(x) = (J-1) + x^{m_s} + \\ (J-1)x^{m_s-1} + x^{m_s} \\ = (J-1) + (J-1)x^{m_s-1} + 2x^{m_s}, \end{aligned}$$

with all coefficients strictly larger than 1. Note that $p_1(x)$ corresponds to the first column of the protograph and $x^{m_s-1}p_2(x)$ to the $2m_s^{\text{th}}$ column. Thus, we have $\langle S \rangle^* = 2m_s$. Since we have $\langle S(J, 2J, m_s) \rangle^* \leq 2m_s$ from Proposition 2, we conclude that $\langle S(J, 2J, m_s) \rangle^* = 2m_s$.